\documentclass[%
 reprint,
Chaplygin,
]{revtex4-2}

\usepackage{graphicx}
\usepackage{dcolumn}
\usepackage{bm}
\usepackage{verbatim}
\usepackage{amsmath}
\usepackage{hyperref}
\usepackage[utf8]{inputenc}
\usepackage[english]{babel}

\begin{document}

\preprint{APS/123-QED}

\title{The growth of fluctuations in Chaplygin gas cosmologies: A nonlinear 
Jeans scale for unified dark matter}


\author{Abdelrahman Abdullah}
 \altaffiliation[Also at ]{Physics Department, Faculty of Science, 
 Cairo University}
\author{Amr A. El-Zant}%
 \email{amr.elzant@bue.edu.eg}

\affiliation{
 Centre for Theoretical Physics, The British University in Egypt,  Sherouk City 11837, Cairo, Egypt
}%
\author{Ali Ellithi}
\affiliation{Physics Department, Faculty of Science, Cairo University,  Cairo, Egypt}%

\date{\today}

\begin{abstract}
Unified dark matter cosmologies economically combine 
missing matter and energy in 
a single fluid. Of these models, the standard Chaplygin gas 
is theoretically motivated, but faces problems in  
explaining large scale structure if 
linear perturbations are directly imposed 
on the homogeneous fluid. 
However, early formation 
of a clustered component of small halos is sufficient (and necessary) 
for hierarchical clustering to proceed in a CDM-like component 
as in the standard scenario,
with the remaining homogeneous component
acting as dark energy.  We examine this possibility. 
A linear analysis shows that a critical Press-Schecter 
threshold for collapse 
can generally only be reached for generalized Chaplygin gas models that mimic $\Lambda$CDM, 
or ones where superluminal sound speeds occur. 
But the standard Chaplygin gas case
turns out to be marginal, with overdensities reaching order one in the linear regime. 
This motivates a nonlinear analysis.  A simple  
infall model suggests that collapse is indeed possible 
for perturbations of order 1~kpc and above; 
for, as opposed to standard  gases, 
pressure forces decrease with increasing densities, allowing for the collapse of linearly stable systems.  
This suggests that a cosmological scenario based on the  standard Chaplygin gas 
may not be ruled out from 
the viewpoint of structure formation, as often assumed. On the other hand, 
a 'nonlinear Jeans scale', constricting growth to scales
$R \gtrsim {\rm kpc}$, which may be relevant 
to the small scale problems of CDM, is predicted.  Finally, 
the background dynamics of clustered Chaplygin gas cosmologies 
is examined and confronted with 
observational datasets. It is found 
to be viable (at 1-sigma), with a mildly larger $H_0$ than $\Lambda$CDM, 
if the clustered fraction  is larger than $90 \%$.       

\end{abstract}

\maketitle

\maketitle


\section{Introduction}

If general relativity holds on galactic and cosmological scales, the 
vast majority of the matter and energy content of our universe 
must be in unknown form. In the standard `double dark' scenario, 
most of the dark sector is composed of dark energy in the form of a cosmological constant ($\Lambda$),  while the rest is in a cold dark matter (CDM) component~\cite{frenkwhite, PrimackR}. 
On the phenomenological level, the
model is highly predictive and successful, 
the tension between the locally measured values of the expansion 
rate and that inferred from the cosmic 
microwave background (CMB)~\cite{Valentino_Silk21} notwithstanding. 
Other issues, include small (galactic) scale 
problems at low redshift~\cite{delpopolo, bullock, SalucciAAR19}, and apparent discrepancies related to unexpectedly  
early galaxy massive galaxy and black hole formation~\cite{DeRosa_SMBH14,Wu_SMBH_Natur15, Dusty_Gal_15, 
Imp_earl16,  First_StarsBH_Volont16, Bouwens_halostellar17, Massive_Q_Glaze17,  Dust_quas17, Silk_LCDM_SMBH18, 
SMBH19, Queis_Obs_mod19, Bromm_SMBH_Rev19, Dominant_Yoshi2019,EZK20}.

On the fundamental level, despite extensive direct and collider searches, the prime CDM candidate, 
the weakly interacting massive particle (WIMP) 
believed  to naturally arise in extensions of the standard model, has not materialized.   
The parameter range open to the 'WIMP miracle' has been thus shrinking, 
at the same time that solutions to the aforementioned small scale problems 
have been invoking alternative dark matter candidates; such as warm dark matter~\cite{Colin2000, Bode2001,  Maccio2012b, Lovell2014, El-Zant2015}, self interacting dark matter~\cite{Spergel2000,Burkert2000,Kochanek2000,Miralda2002,
Zavala2013, Elbert2015}, dissipative dark matter~\cite{Randall13, Randall15, Vagnozzi15, Vagnozzi16}, 
and fuzzy dark matter made of ultra light axions~\cite{Goodman2000, Hu2000, Schive2014, Marsh2014, Hui_etal2017}). Dark energy, on the other hand, remains evermore elusive. 
Particularly perplexing is its  small magnitude, 
and its relatively recent domination of the 
cosmic energy budget. These issues have been the subject of intense investigation, seeking alternative to the cosmological constant as drivers of late cosmic acceleration, including modifications of fundamental gravitational law
(e.g.\cite{Li_2011,2012IJMPD..2130002Y,DERev18} for reviews).

Dark energy effectively contributes a negative pressure term in the Friedmann acceleration equation,
thus accelerating the late background dynamics. 
Earlier on, when the universe is denser, 
the dark energy contribution is  small; with radiation, then pressureless matter dominating. 
Matter also dominates in the late universe on nonlinear scales, of galaxies and clusters. which are likewise characterized by relatively high density and little pressure. 
It is therefore natural to ask whether the 'double dark' sector is simply composed of single component; one with negative pressure at low density and essentially zero pressure at high densities.

The most extensively studied among  unified dark matter models that fit the bill  
are generalized Chaplygin gas cosmologies~\cite{2001PhLB..511..265K, 2002PhRvD..66d3507B}, 
where the  associated fluid comes with 
an equation of state relating the pressure and densities 
\begin{equation}
p = - \frac{A c^2}{\rho ^\alpha }, 
\label{eq:state}
\end{equation}
and where $A$ and $\alpha$ are parameters that may be determined empirically
to fit observations. Attractive aspects of such models include the simplicity via which the dark sector is unified through Eq.~(\ref{eq:state}); 
their relevance to some problems affecting the standard model, 
such as the $H_0$ tension, is also of interest~\cite{VagnozziChap19}.  
The original Chaplygin gas  model, with $\alpha = 1$, is particularly well motivated from a theoretical point of view~\cite{jackiw2000particle, gorini2005chaplygin}. 
However, in its context,  large scale 
perturbations, growing relatively late on a homogeneous background,
are affected by the increasing sound speed in the expanding medium. 
The pressure forces propagation rate thus catches 
up with the rate of gravitational collapse and halts the condensation.
The perturbations thus become 
Jeans stable; oscillating acoustically and (Hubble) damping instead of growing. This imprints 
strong unobserved signatures on the matter density fluctuation power spectrum. The predictions regarding structure formation are, in this context, in catastrophic tension with observations~\cite{2004PhRvD..69l3524S}). 

Including a baryonic component improves the situation~\cite{2003PhRvD..67j1301B}. 
Nevertheless, values of $\alpha$  very close to zero (and therefore to the standard model) 
or greater than one (when the sound speed may become superluminal),
are still favoured~\cite{2008PhRvD..78j3523F, 2008JCAP...02..016G, 2010PhRvD..81h7303F}). 
Indeed, combinations of supernovae data, CMB and baryon acoustic oscillations data, appear 
to conclusively constrain  Chaplygin gas cosmologies to the neighborhood of $\Lambda$CDM~\cite{2010PhRvD..81f3532P, 2013PhRvD..87h3503W}). Non-adiabatic perturbations (e.g. \cite{2003PhRvD..68f1302R, 2005CQGra..22.4311Z,2007JPhA...40.6877B}) were also considered. It is not clear however whether the growth of such perturbations would remain impervious to pressure forces into the non-linear regime. 

One may immediately remark, however, that the above constraints are inferred 
by considering a nearly homogeneous Chaplygin gas that remains unclustered 
on all scales, including the smallest  ones (corresponding to dwarf galaxy halos and smaller).  
For example, Sandvik et. al. \cite{2004PhRvD..69l3524S} examine linear perturbations in a homogeneous Chaplygin gas fluid on scales corresponding 
to comoving wavenumbers $k < 1~{\rm Mpc^{-1}}$.  
But in a hierarchical scenario,  the medium 
in which such perturbations are probed  should not be considered homogeneous. 
Indeed, in the standard CDM pictures it  has already clustered 
on a hierarchy of smaller scales; 
from the smallest earth mass gravitationally  bound structures up to galactic and cluster scale halos.

In a successful unified dark matter scenario 
large scale perturbations would occur in an already clustered medium. Such a medium   would act as pressureless CDM, readily allowing further clustering. 
The hierarchically forming halos would host galaxies with spatial distribution  
that may be expected to be quite close to the standard CDM case, with commensurately similar linear matter power spectrum.  Furthermore, if clustering is efficient, simple 
arguments suggest that the background evolution  can also
be rendered compatible for observations, 
even for the theoretically attractive 
case of $\alpha = 1$~\cite{2014PhRvD..89j3004A}.

The central question is therefore not whether the evolution of large scale (and thus late growing)  
perturbations in a homogeneous unified dark matter 
fluid is compatible with data, but whether such a medium can sufficiently cluster early on.
For this would enable it to exhibit a significant CDM like component, which can 
hierarchically condense into halos and hence host galaxies. 
Answering this in full requires, in principle, 
detailed modelling of the nonlinear collapse and clustering process. 
This has not 
been available, due to the novelty and presumed complexity of modelling the hydrodynamics of fluids with anomalous equations of state, involving 
negative pressure with absolute value decreasing with density. 
But this anomalous form itself suggests that examining the nonlinear regime must be considered 
central to any investigation; as, in a Chaplygin gas, pressure forces become less significant
with increasing density, a linearly (Jeans) stable system may be nonlinearly unstable against 
gravitational collapse, providing for a medium of collapsed halos that may cluster hierarchically. 
This is in stark contrast to a regular gas, where the sound speed and pressure forces necessarily increase with density, which in turn implies that a  Jeans stable gas generally remains stable against nonlinear perturbations. 
Pending  full hydrodynamic treatment,  estimates 
of the effect of nonlinearity, circumventing complications of unknown hydrodynamics 
while capturing the essentials, thus seem crucial to adequately 
evaluating the efficacy of Chaplygin 
gas cosmology. 

Here we revisit the linear and nonlinear stability of generalized Chaplygin
gases, with the aforementioned remarks in mind. We confirm previous results 
suggesting that linear clustering is most efficient for either very small
values of $\alpha$ in Eq.~(\ref{eq:state}), or larger ones associated with
superluminal sound speeds (Section~\ref{sec:level2}). Nevertheless,  we point out 
that even a linear analysis 
predicts that perturbations can come tantalizingly close (at small scales) to the critical  
(Press-Schecter based value) for collapse in the fiducial case of standard Chaplygin gas with $\alpha = 1$. This further motivates examination of the nonlinear regime. 
In Section~\ref{sec:nonlinear} we review 
some previous attempts at such an analysis, before proceeding  
to estimate the importance of pressure forces along an 
inhomogeneous spherical collapse model, 
avoiding complications that come with shell crossing and possible formation of shocks.
In Section~\ref{sec:bback} we discuss the basic characteristics and viability of the background evolution of a clustering Chaplygin gas cosmology. 

\section{\label{sec:level2} Perturbations in the linear regime}

We briefly review the evolution of the homogeneous generalized 
Chaplygin gas and the fixing of its parameters. We then describe the effect of 
perturbations and present the resulting matter power spectra and 
associated RMS mass fluctuations on various scales. 

\subsection{Background evolution of homogeneous Chaplygin gases}
\label{sec:parameters}

By inserting the equation of state~(\ref{eq:state}) into the energy conservation law
\begin{equation} 
\dot{\rho} + 3 H (\rho + p/c^2) = 0,  
\end{equation}
where $H$ is the Hubble parameter,  one obtains the evolution equation for the homogeneous 
generalized cosmological Chaplygin gas:
\begin{equation} 
\rho(t) = \bigg[ A + \frac{B}{a^{3(1+\alpha)}}\bigg]^{\frac{1}{1+\alpha}}. 
\label{eq:density}
\end{equation}
Here, 
$a(t)$ is the scale factor and $A$ and $B$ are constants that can be related to the dark energy and matter cosmic contents respectively:
\begin{equation} 
A = \rho_c^{1+\alpha} \left(1 - \Omega_m^{1+\alpha} \right), \:\:\:
B = \rho_c^{1+\alpha}  \Omega_m^{1+\alpha}, 
\label{eq:ABH}
\end{equation}
where $\rho_c$ is the current total density, which is thus assumed to correspond to the 
critical closure density, and $\Omega_m = \rho_m/\rho_c$ is the current ratio of matter to total density.~\footnote{This ignores the baryon contribution, as the whole matter density is supposed 
to emanate from the unified dark matter component.
The effect of baryons generally alleviates the problems addressed 
here~\cite{2003PhRvD..67j1301B}; any solution to these problems that is 
effective in their absence should remain valid when they are added.} 
For sufficiently small $a$, the second term is dominant  in ~(\ref{eq:density}), and behaves as $\rho \sim 1/a^3$, in correspondence to the standard matter domination era. On the other hand, the current dark energy dominated era is characterized by
\begin{equation} 
\rho(a=1) = \rho_c = (A+B)^{\frac{1}{1+\alpha}},
\end{equation}
while $\rho \rightarrow A^{\frac{1}{1 + \alpha}}$ specifies the dark energy content.
In turn, for a given $\alpha$, $A$ and $B$ may be fixed by specifying the current contribution of the 
dark energy and matter components to the total energy density. 

\subsection{Growth and oscillation of linear perturbations}

The relativistic equation that describes the growth of small matter 
overdensity perturbation modes in a nearly homogeneous fluid may be written as 
\begin{align}
\begin{split}
\ddot{\delta}_k+ H \dot{\delta}_k [2 - 3(2w-c_s^2)] - \frac{3}{2}H^2\delta_k[1-6c_s^2 \\ +8w - 3w^2] = -\left( \frac{kc_s}{a} \right)^2 \delta_k,
\end{split}
\label{eq:perturb}
\end{align}
where the equation of state parameter 
$w$ and the sound speed $c_s$ are given by
\begin{equation} 
w \equiv \frac{p}{\rho} = - \bigg[1 + \frac{\Omega_m^*}{1 - \Omega_m^*} a^{-3(1+\alpha)} \bigg   ]^{-1}, \end{equation}
and
\begin{equation} 
c_s^2 \equiv \frac{\partial p}{\partial \rho} = -\alpha w c^2,
\label{eq:SSB}
\end{equation}
with $\Omega_m^* = \Omega_m^{1+\alpha}$ and $\Omega_m$ the current matter density 
parameter~\cite{2004PhRvD..69l3524S}. In Eq.~(\ref{eq:perturb}) $c_s$ is assumed to be expressed 
in units of $c=1$.

Equation~(\ref{eq:perturb}) is solved numerically by changing
the independent variable from $t$ to $\ln a$. Accordingly,
\begin{equation} 
\frac{d}{dt} = H \frac{d}{d\ln a},  
\end{equation}
and
\begin{equation} 
\dot{\delta}_k =  H^2 \delta{''} + \frac{1}{2} (H^2)' \delta', 
\end{equation}
where $' \equiv \frac{d}{d \ln a} $ and
\begin{equation} 
\xi \equiv \frac{(H^2)'}{2H^2} = -\frac{3}{2} \left(1+\left(\frac{1}{\Omega_M*}-1\right)a^{3(1+\alpha)} \right)^{-1}. 
\end{equation}
Equation~(\ref{eq:perturb}) can then be written as
\begin{align}
\begin{split}
\delta^{''}_k + [2 + \xi -3(2w - c_s^2)] 
\delta^{'}_k 
=\\
\bigg[
\frac{3}{2} \left(1 - 6c_s^2  
+ 8 w -3 w^2 \right) 
-  \left(\frac{k c_s}{aH}\right)^2 
\bigg] \delta_k.
\end{split}
\label{eq:linpertscaled}
\end{align}

We immediately note the appearance, in this context, 
of a (non-relativistic) Jeans length, arising from the 
presence of the pressure term, 
\begin{equation} 
\lambda_J = \sqrt{\frac{\pi |c_s^2|}{G\rho}}.
\label{eq:Jeans}
\end{equation}
As a result if, as will be assumed here, $c_s^2 > 0$ ($\alpha >0$), 
linear perturbations on scales 
below $\lambda_J$ will acoustically oscillate, and damp with the expansion,
rather than grow.  
(If $c_s^2 < 0$. the perturbations below 
the critical scale will grow exponentially~\cite{2004PhRvD..69l3524S,2010PhRvD..81f3532P}).

The Jeans length, as usual, reflects a competition between the sound crossing time and  
the gravitational collapse time, determined by the dynamical time $\approx  (G \rho)^{-1/2}$. 
In the matter dominated regime $\rho \propto 1/a^3$ and $c_s^2 \propto \alpha/\rho^{\alpha+1}$. For 
$\alpha > 0$ ($c_s^2 > 0$), $\lambda_J$ is a real number that rapidly  
increases as 
\begin{equation}
\lambda_J \propto \sqrt{\alpha} a^{3 (\alpha + 2)/2}
\label{eq:Jprop}
\end{equation}
as the universe expands. For $\alpha = 1$, for example, this corresponds to 
a steep increase $\sim a^{9/2}$. 
Thus according to such estimates --- derived 
on the assumption that the perturbed medium remains nearly homogeneous on scales smaller than $\lambda_J$ ---
the growth of late forming structures is suppressed, unless $\alpha$ is very small; 
as, otherwise, the  rate of gravitational collapse cannot catch up with the swiftly increasing sound
horizon. 

In this context, the large scale structure power spectrum  will
be dominated by acoustic oscillations, if perturbations on the relevant 
scales are evolved to the present while imprinted on a homogeneous background.
If smaller scales have collapsed into halos, however, the situation is qualitatively 
different. For, in this case, a  clustered CDM-like medium forms,  composed of 
halos that may merge into larger structures, as in the standard scenario. 
This merging process, also as in the standard scenario, reflects the 
growth of larger scale perturbations, which are now imprinted on the 
clustered medium and not on the initial homogeneous background. 
As the clustered medium is made of high density halos, and is therefore essentially
pressureless, it should act as a standard dark matter component,  and thus clusters 
on larger scales as CDM does. 
The rest of this study aims at investigating this possibility. 

\subsection{Power spectrum and RMS fluctuations}

We start with a standard Gaussian random field of density perturbations. It is obtained 
using the publicly available code CAMB~\cite{lewis}, with 
cosmological parameters $h = 0.69$, $n_s = 1$, $\Omega_{\Lambda} = 0.71$, $\Omega_m = 0.29$, $\Omega_k= 0.001$. This fixes our initial conditions at 
$z = 100$, when the Chaplygin gas pressure is still entirely negligible, and 
the cosmological model and its spectrum of fluctuations are
indistinguishable from $\Lambda$CDM. 

We then evolve realizations of this Gaussian random field  
by integrating Eq.~(\ref{eq:linpertscaled}), using an adaptive Runge-Kutta method, 
to obtain the matter power spectrum
\begin{equation} 
P(k) = \langle |\delta(k)|^2 \rangle 
\end{equation}
at different redshifts.
and also the associated dimensionless power spectrum,
\begin{equation} 
\Delta ^2 (k) \equiv \frac{1}{2 \pi ^2}  k^3 P(k), 
\label{eq:dimenps}
\end{equation}
which measures the contribution of perturbations per unit logarithmic interval at wave number 
$k$ to the variance of matter density fluctuations. This variance may in turn be given 
in terms of the power spectrum, filtered on spatial scales $R$, such that 
\begin{equation} 
\sigma_M^2(R) = \frac{1}{2\pi^2} \int P(k) W_R^2(k)k^2 d k,  
\end{equation}
where $W_R(k)$ is the Fourier transform of the spatial filter (henceforth assumed top hat).

\subsection{Results: marginal stability at $\alpha =1$}
\label{sec:mar_stab}

\begin{figure}[!htbp]                           
\centering
\includegraphics[width=0.53\textwidth]{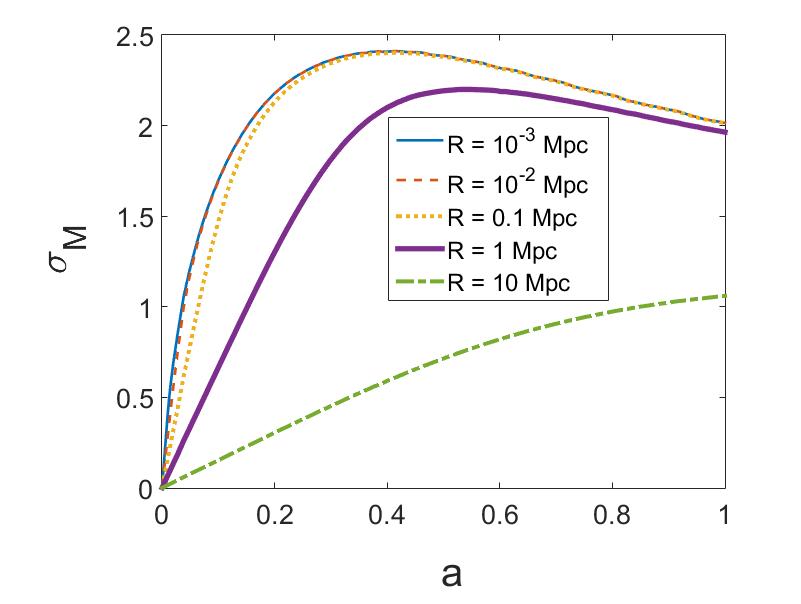}
\caption{The evolution of the mass dispersion $\sigma_M (R, a)$, 
with scale factor $a$, 
for a generalized 
Chaplygin gas with $\alpha = 10^{-5}$ (in Eq.~\ref{eq:state}).   
smoothed on the indicated (comoving) scales $R$.}
\label{fig:SigmaM_A1e-5}
\end{figure}

As expected from previous work, discussed in the introduction, when  
the index $\alpha$ in~(\ref{eq:state}) is small, the growth 
of perturbations in a homogeneous generalized Chaplygin gas 
is closest to the standard case involving CDM. 
Fig.~(\ref{fig:SigmaM_A1e-5}) shows the RMS 
dispersion $\sigma_M (R)$ 
for the Chaplygin fluid for  $\alpha = 10^{-5}$.
Perturbations at all smaller scales pass the critical Press-Schecter threshold
of about $\sigma_M = 1.7$ (e.g., \cite{Peacock}), 
the Jeans length catching up with the scale of the collapsing object only 
well into the nonlinear regime, when bound halos are expected to have already formed. 
The largest perturbation scale shown ($R= 10 \: {\rm Mpc}$) remains largely unaffected by the
generalized Chaplygin gas pressure. 
The effect of pressure forces in this case (of small $\alpha$) 
is small due to the smallness of the prefactor $\alpha$ in the 
expression of the sound speed
Eq.~(\ref{eq:SSB}). 

\begin{figure}[!htbp]
\centering
\includegraphics[width=0.53\textwidth]{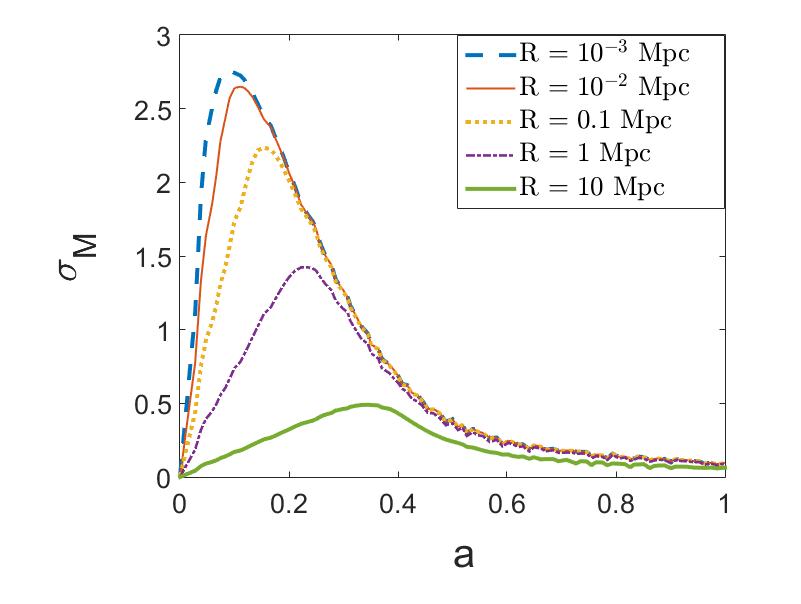}
\caption{Mass dispersion $\sigma_M (R, a)$,  when $\alpha = 2$ in Eq.~(\ref{eq:state}).}
\label{fig:SigmaM_A2}
\end{figure}

For large values of $\alpha$,  the suppression of structure formation 
is also alleviated. Here,  the reduction is  
due to a steep variation of the sound sound with density. 
The results for $\alpha = 2$ are shown in Fig.~\ref{fig:SigmaM_A2}.
Note that here the Press-Schecter threshold of around 1.7 is only 
passed on scales smaller than 1~Mpc, 
as the steep dependence of the pressure on density 
enables the growth at higher redshift when the background density 
is relatively large. This is the epoch where smaller perturbations grow most, due to their  earlier headstart (resulting from the shape of the standard initial dimensionless 
power spectrum). The growth of larger fluctuations, which 
takes place later, is highly suppressed (as they evolve later, when the 
background density is lower, and Jeans 
scale growing as in equation Eq.~\ref{eq:Jprop}). 
The collapse of the smaller perturbations is nonetheless 
sufficient to allow for a pressureless clustered CDM-like component, 
from which higher scales can hierarchically cluster.
However,  the sound speed in the a remaining homogeneous 
background fluid can become superluminal at lower redshift, 
Whether this is physically admissible has been discussed in~\cite{2008JCAP...02..016G}.

\begin{figure}[!htbp]
\centering
\includegraphics[width=0.53\textwidth]{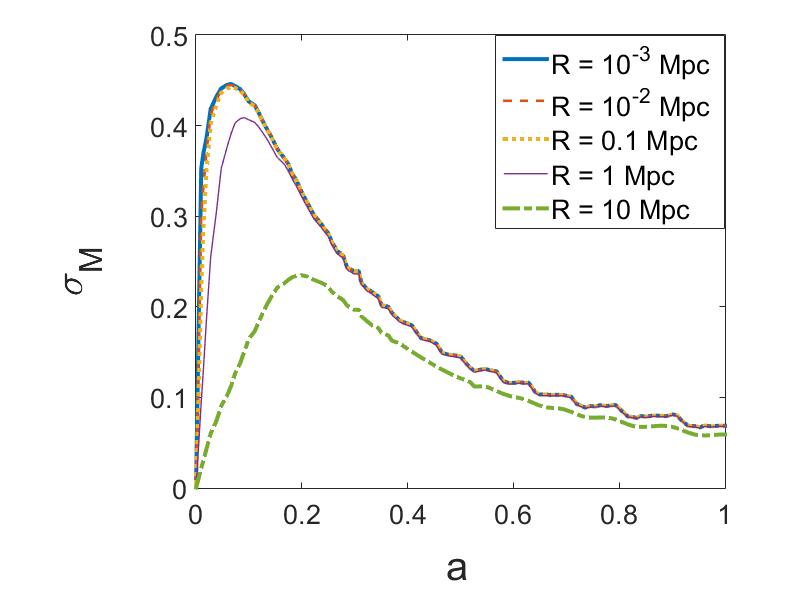}
\caption{Mass dispersion $\sigma_M (R, a)$ when $\alpha = 10^{-2}$ 
in Eq.~(\ref{eq:state})}
\label{fig:SigmaM_A1e-2}
\end{figure}

Intermediate values of $\alpha$ correspond to situations whereby neither a steep decrease 
in sound speed with density nor small prefactor $\alpha$ in Eq.~(\ref{eq:SSB}) may 
sufficiently suppress the pressure forces. Fig.~\ref{fig:SigmaM_A1e-2} shows the case 
$\alpha = 10^{-2}$, 
where all perturbations acoustically oscillate and damp well before 
reaching the critical Press-Schecter threshold of $\sigma_M = 1.7$.
In this case, therefore, no structure is expected to form on any scale. 

Of particular interest is the fiducial case of 
$\alpha = 1$. Here, the Press-Schecter threshold is not formally reached
but the  necessary condition for nonlinear growth, 
$\sigma_M \approx 1$, 
is attained on the smaller scales shown, of $R \le 0.1 {\rm Mpc}$. 
\begin{figure}[!htbp]
\centering
\includegraphics[width=0.53\textwidth]{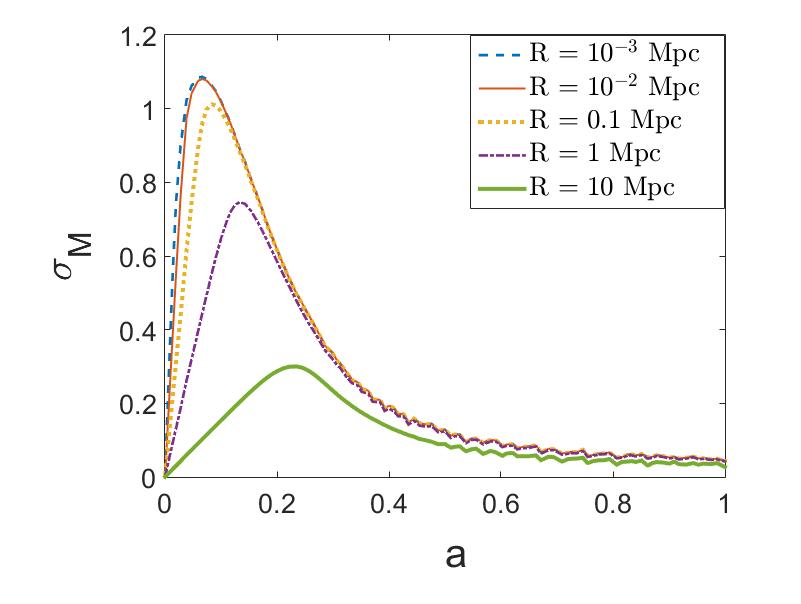}
\caption{Mass dispersion $\sigma_M (R, a)$ for the case of the standard Chaplygin gas, with $\alpha = 1$ 
in Eq.~(\ref{eq:state}).}
\label{fig:SigmaM_A1}
\end{figure}
This 
leaves open the possibility of collapse in the nonlinear regime, which is characterized by increased 
density and steeply decreased pressure. 
And with this the emergence of a CDM-like component composed 
of collapsed halos, and 
the possibility of 
hierarchical clustering thereof.

\begin{figure}[!htbp]               
\centering
\includegraphics[width=0.51\textwidth]{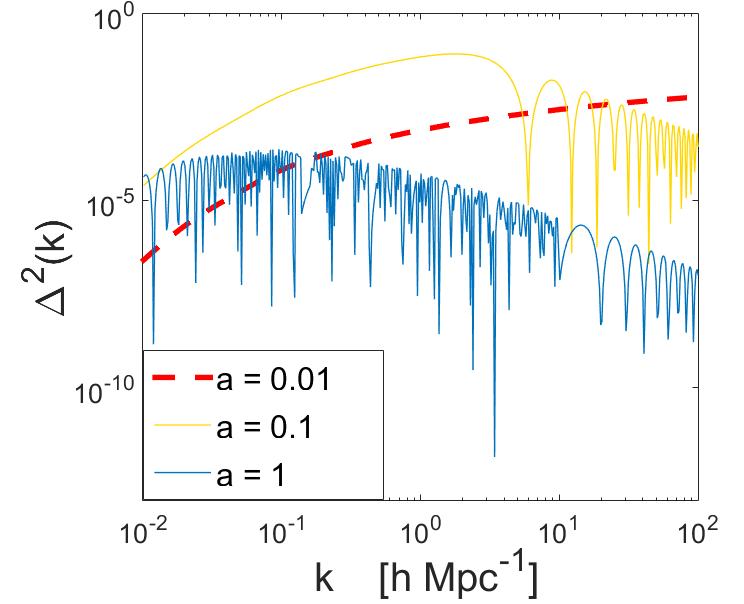}
\caption{The evolution, with scale factor,  of the dimensionless power spectrum 
of linear perturbations in a standard Chaplygin fluid ($\alpha = 1$) cosmology.}
\label{image-DimensionlessPS}
\end{figure}

Further clarification of a possible route to hierarchical structure formation
in such a situation (i.e., for the case  $\alpha = 1$), may be illustrated by explicitly examining the 
evolution of the dimensionless 
power spectrum. This, we plot in Fig.~\ref{image-DimensionlessPS}.  
As  is clear, linear perturbations in the homogeneous fiducial Chaplygin gas 
universe at $z=0$ ($a = 1$) are 
eventually acoustically suppressed at all scales. Nevertheless, as one goes back to earlier times (larger background $\rho$), significant growth can occur,
before oscillations and expansion damping dominate. This gives rise to the 
peaks in the dispersion $\sigma_M$, which touch the nonlinear regime and approach 
the Press-Schecter collapse threshold  in 
Fig.~\ref{fig:SigmaM_A1}. 

The marginal stability observed in the linear regime may have crucial consequences 
for the clustering of the Chaplygin gas. 
For the fiducial Chaplygin gas has the peculiar property that the  
pressure forces in general steeply decrease in magnitude with density: 
 $\sim c_s^2/\rho \propto \sim 1/\rho^3$. Therefore, if collapse is so nearly 
reached in the linear context, it  may very well actually occur 
if the nonlinear increase in density is adequately 
taken into account. This is what we attempt to 
investigate in the next section, by invoking a simple spherical infall model.

\section{Nonlinear regime: 
expected clustering for $\alpha =1$ and minimal mass scale}
\label{sec:nonlinear}

\subsection{Motivation}

\begin{figure}[!htbp]
\centering
\includegraphics[width=0.51\textwidth]{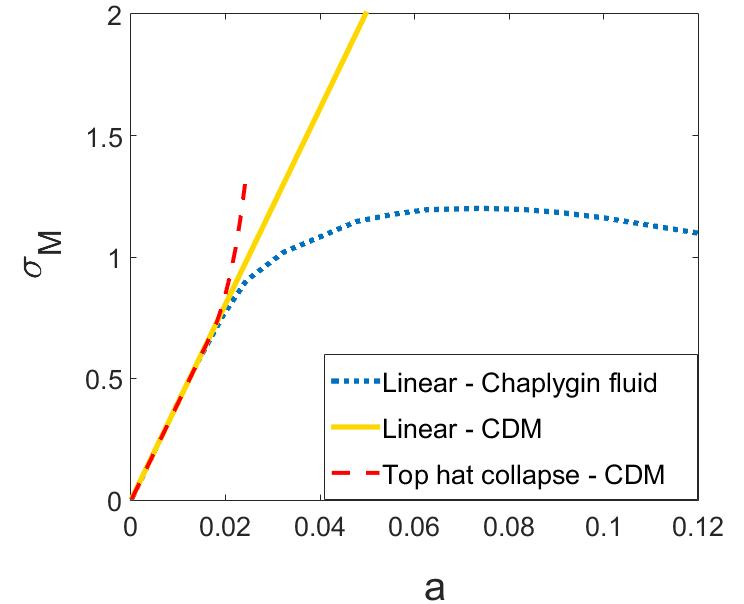}
\caption{The evolution, in the linear regime,
of the mass dispersion $\sigma_M (R, a)$ on 1 kpc scale in 
the standard Chaplygin fluid, compared with the 
evolution of overdensity corresponding CDM case and its nonlinear (homogeneous) 
top hat counterpart.}
\label{image-SigmaM_3}
\end{figure}

In Fig.~\ref{image-SigmaM_3}
we compare the linear growth of perturbations, on a kpc scale, 
in a pressureless CDM-like fluid  to that in the fiducial Chaplygin gas ($\alpha = 1$), 
and to the nonlinear growth in a homogeneous top hat 
collapse model of a pressureless medium~(e.g.~\cite{Peacock}).   
As is clear, the growths are similar up to $\sigma_M \approx 1$. Beyond that, as may be expected, the overdensity associated with nonlinear 
top hat collapse diverges from the linear 
one in the pressureless case. 
In the linear analysis, the overdensity 
growth in the Chaplygin gas case is then reversed, 
as the pressure term overtakes self gravity in significance. 
Again, if one was dealing with a 
regular laboratory gas, this trend would be expected reinforced in the nonlinear regime,  
since the sound speed and associated  
pressure forces increase as the densities increase. 
At an overdensity of order one however, the squared sound speed in 
a Chaplygin gas $c_s^2 = d p/d r \sim 1/\rho^2$ has already decreased by a factor of 4. 
This implies (for a given density gradient) 
an order of magnitude (a factor of 8 to be precise)
decline in  pressure forces 
$\frac{1}{\rho} d p/d r \sim \frac{c_s^2}{\rho}$. This is not taken into account in a linear
analysis, which effectively assumes 
that the pressure inside the overdensity can be characterised by the 
steeply rising sound speed of the rapidly decreasing density of the expanding background.
(Indeed, with $c_s^2/\rho \sim 1/a^9$!).  

Further compression implies additional 
suppression of pressure forces, relative to the linear regime, 
while the competing gravitational forces are enhanced. 
Once turnaround is achieved, 
the pressure forces can only become less important, as the density now increases 
in absolute terms (and not only relative to the expanding background). 
It would therefore seem entirely plausible that a nonlinear 
analysis would predict collapse instead of re-expansion of small scale 
perturbations. Allowing for halos that can hierarchically cluster as in the standard, CDM-based, 
scenario.

Given this, remarkably little work regarding the nonlinear stability of 
the Chaplygin gas has been carried out. Here, we briefly discuss a couple of examples, 
pointing out that the conclusions attained from the simplified treatments 
are far from conclusive. 
Given the complexity of the problem at hand, 
and the largely unexplored hydrodynamics of the peculiar equation 
of state involved, 
this is not surprising. 

Bili{\'c} et al.~\cite{tupper} have used the continuity and Euler-Poisson system of 
equations in an expanding universe
to derive an equation that describes the growth of overdensity perturbations
\begin{align}
\begin{split}
a^2 \delta'' + \frac{3}{2} a \delta' -\frac{3}{2} \delta (1 + \delta) - \frac{4}{3} \frac{(a \delta')^2}{1+\delta} \\ - \frac{1+ \delta}{a^2 H^2} \frac{\partial}{\partial x_i} \left( \frac{c^2_s}{1+ \delta} \frac{\partial \delta}{\partial x_i} \right) =  0,
\end{split}
\label{eq:Tupp}
\end{align}
where a $'$ denotes the derivative with respect to scale factor $a$, 
$\mathcal{H}$ is a local Hubble parameter (describing the expansion, or contraction, rate  
of each shell inside a spherical overdensity), and $x$ is a radial variable.  
In order to study the nonlinear growth,  a self similar solution, 
involving a time dependent scale $R$, such that 
$\delta (a,\bm{x}) = \delta_R(a) f(x/R)$, 
was proposed. 
It is not clear however, that such a solution exists, and for which 
form of $f$. This is crucial, as in the presence of significant 
pressure gradients, such a form may very well not in fact exist
(even with gravity alone acting, they generally exist only for power 
law initial conditions~\cite{Fill_Gold}). 
In the aforementioned work, a Gaussian 
$f(x/R) = \exp(\frac{-x^2}{2R^2})$ was {\it a priori} assumed, 
and the equation arising from  substituting the resulting
$\delta (a,\bm{x})$ into (\ref{eq:Tupp})
still involved the Gaussian and its derivatives. 
To get rid of it, the resulting equation was 
arbitrarily solved at $x = 0$. However, 
transforming a partial 
differential equation, describing fluid flow (and its continuity),  
into an ordinary one at an arbitrary point cannot be considered 
generically valid.  It also appears that the background sound speed, 
and not the much smaller one inside the perturbation (as discussed above), 
was used. 
In turn, the principal conclusion
from that work, which points towards inefficient Chaplygin 
gas clustering, even in the 
nonlinear regime, seems to lack sufficient foundation
to rule out the viability of the otherwise appealing 
unified dark matter cosmologies based on the standard Chaplygin gas.

Another attempt at examining nonlinear growth in Chaplygin gases was undertaken  
by Fernandes et al.~\cite{fernandes2012spherical}. Here, a top hat collapse model was invoked, 
in such a way that 
a homogeneous density profile was assumed 
inside a spherical perturbation. with a pressure discontinuity 
at the boundary, separating the overdense region from the homogeneous background. 
This thus ignores pressure gradients inside the perturbation and concentrates 
all pressure forces at the boundary. 
The results are also anomalous --- and in disagreement with linear analysis --- 
in the sense that they suggest that pressure forces associated with 
larger positive values of $\alpha$ tend to actually
speed up the collapse relative to the 
pressureless case of $\alpha =0$. 

In the following, we describe a model that may help estimate 
the possibility of collapse and formation of self-gravitating halos in
standard Chaplygin gas cosmologies.  The principal goal is to take into 
account, in simplest terms, the effect of peculiar phenomenon of the decrease in the magnitude
of the pressure with the increasing (over)density, characteristic of any unified dark matter 
fluid.

\subsection{Possibility and scale of unhindered collapse in a spherical infall model}

\subsubsection{Basic idea}
\label{sec:basic_id}

To our knowledge, the hydrodynamics of a negative pressure gas has not been explored in any detail. In the mathematical literature, it is known that solutions of the Riemann problem lead to shocks \cite{2009ArRMA.191..539S, Wang2013TheRP, YANG20125951}, which may
accompany eventual shell 
crossing in a nonlinear collapse model. In the following 
we wish to circumvent such complications, while obtaining a reasonably realistic estimate for the possibility 
of gravity overcoming pressure forces, so as to allow for 
collapse in the standard Chaplygin gas.

For this purpose we use a simple spherical infall model, where, in  the absence 
of pressure forces, all shells reach their 
maximum radius and turnaround before any shell crossing occurs.  
We solve the  dynamics of the model
and estimate the pressure forces along the unperturbed motion. 
Given that Chaplygin gas pressure forces necessarily become less important with increasing density, 
while gravitational forces become more potent,  the maximum strength of the 
pressure forces to gravitational force, at every shell, will occur close to its 
maximum expansion at turnaround.  In this context, if we can show that pressure forces 
are small compared to gravitational ones before and around the 
turnaround, for all shells, 
then this suggests they are  negligible throughout the evolution, and thus collapse
akin to that in pressureless CDM may occur. 

\subsubsection{Model}
\label{sec:model}

In the context just set, we consider the dynamics of a pressureless matter 
perturbation in a flat, matter dominated universe. As we are interested in the early collapse of the smallest structures, 
when dark energy is not important, we limit ourselves to the case of an Einstein de Sitter universe. 
Without shell crossing, 
the mass $M = M (< r)$, within a shell at radius $r$ 
of a spherical perturbation, is conserved. The dynamics of its 
Lagrangian radius 
is then simply determined by $\ddot{r} = - G M/ r^2$, with (specific) energy integral 
$\frac{1}{2} \left(\frac{d r}{d t}\right)^2 - \frac{G M}{r} = E$, and parametric solutions
in terms of a phase angle $0 \le \theta \le 2 \pi$: 
\begin{eqnarray}
\nonumber
r = A_{\rm sh}  (1 - \cos \theta)\\
t = B_{\rm sh}  (\theta - \sin \theta),
\label{eq:cycloid}
\end{eqnarray}
where $\rm A_{\rm sh} = G M/ |E|$ and $\rm A_{\rm sh}^3 = G M B_{\rm sh}^2$, 
and the subscripts indicate that these constants are specific to each shell. 
They can be fixed by conditions at time $t_i$, where the linear 
regime may be assumed to reign. The initial velocity of  a shell corresponds
to the cosmological expansion determined by the Hubble parameter $H_i$, 
minus a peculiar (inward) velocity term  $\bar{\delta_i}(r_i) H_i r_i/3$. The 
initial average overdensity within radius $r_i$ is given by  
$\bar{\delta}_i (r_i) + 1 = M /  \bar{M}_i$. 
Using the energy integral, one finds $A_{\rm sh} = \frac{3}{10} \frac{r_i}{\bar{\delta} (r_i)}$
and $B_{\rm sh} =\frac{1}{2H_{0}} \left( \frac{5}{3} \frac{\bar{\delta_i}(r_i)}{a_i} \right)^{-\frac{3}{2}}$ (e.g., \cite{Peacock, GunnGott72, Loeb06, rubin2013}).

Solutions for the Lagrangian radius $r$ are self similar, in the sense that 
$r/r_i$ depends only on the development angle $\theta$ for any initial radius 
$r_i$ at $t_i$, but $\theta_i$ is different for each shell.  
In accordance with the discussion of the previous subsection, we 
need these (no shell crossing) solutions to be valid at least 
until all shells have achieved maximal expansion 
and turned around. 
In a realistic initial profile, with density 
decreasing with radius, shell crossing will 
first occur at the inner shells. 
To eliminate such crossing (including the innermost shell crossing itself before 
re-expanding) until all shells have turned around, 
one may thus require that $\theta (r_i = 0) \le 2 \pi$, 
when the outermost shell, initially at radius $R_i$,
is turning around; i.e., when $\theta (R_i) = \pi$.  
In general, from the second of equations~(\ref{eq:cycloid}),
a shell initially at dimensionless radius $x = r_i/R_i$ will 
have a development angle 
\begin{equation} 
\theta_t (r_i) - \sin \theta_t (r_i)  = \pi \bigg[\frac{\bar{\delta}_i(x)}{\bar{\delta}_i(R_i)} \bigg]^{3/2}
\label{eq:cylcC}
\end{equation}
at the turnaround of the outer shell. 
This implies that the maximum density contrast allowed between 
the initial central and average density  
is $[\frac{{\delta}_i(0)}{\bar{\delta}_i(R_i)}]_{\rm max} = 2^{2/3}$.

The overdensity within the maximal initial radius $R_i$ can be assumed
to correspond to the RMS dispersion in the linear density 
field $\sigma_M (R_i, t_i)$. To complete the model we then need a profile for the
initial overdensity, in order to determine $\bar{\delta}_i (r_i)$. 
We follow~\cite{rubin2013}
in using a simple generic density profile, with steepness 
adjusted through a parameter $\beta$
\begin{equation}    
\frac{\bar{\delta}_i(x)}{\bar{\delta}_i(R_i)} =  \frac{\delta_i(0)}{\bar{\delta}_i(R_i)} \left(1 - c x^{\beta}\right).   
\label{eq:denprof}
\end{equation}
The corresponding local (as opposed to volume averaged)  
density distribution has the same form but with 
$c \rightarrow c(\beta + 3)/3$. Fig~1 in~\cite{rubin2013} shows the profiles 
for various values of $\beta$ (their Fig.~3 also shows the much 
steeper forms of these density contrast, when evolved till the 
turnaround of the outer shell).
Here we will use three values of $\beta$, reflecting qualitatively 
different behaviors: $\beta=7$, 
corresponding to nearly flat profile in the inner regions and steep decrease in outer 
($x \gtrsim 0.5$) ones; $\beta = 1$, with density decreasing linearly with $x$; and 
$\beta = 0.1$, which corresponds to nearly flat profile in outer region with rapid
increase at small radii. Realistic initial peak profiles, obtained by 
smoothing the linear random field, are consistent with such a flat distribution 
at large radii, with steeper increase in the central region, as Fig.~10 of~~\cite{rubin2013} illustrates.
It also suggests that, for small mass halos collapsing at higher redshifts, which is our prime interest here, 
our relatively small initial maximal average overdensity contrast, chosen to ensure  that no shell crossing occurs, is not unrealistic.

\begin{figure}[!htbp]
\centering
\includegraphics[width=0.49\textwidth]{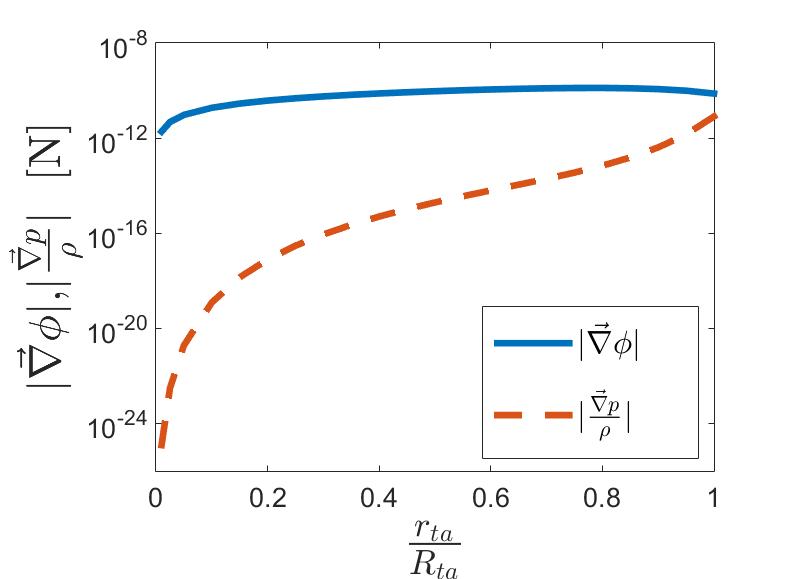}
\caption{Gravitational and pressure forces at turnaround
for initial density profiles given by (\ref{eq:denprof}), with  
$\beta = 7$ (that is flat near centre and steeply 
decreasing for $x \gtrsim 0.5$),  
and $R_i = 1~{\rm kpc}$ (comoving). The system is started 
at $z = 300$ and evolved using (\ref{eq:cycloid}), with initial 
average overdensity inside $R_i$ corresponding to the
RMS fluctuation in the linear field $\sigma_M (1 {\rm kpc})$.
The scaled radius on the $x$-axis denotes the radius of turnaround 
of a certain shell relative to the radius of turnaround of the outer shell. 
As there is no shell crossing up to turnaround of all shells, smaller radii 
correspond to initially smaller $x = r_i/R_i$.}
\label{Beta_7_kpc.jpg}
\end{figure}

\begin{figure*}[!htbp]
\centering
\includegraphics[width=1\textwidth]{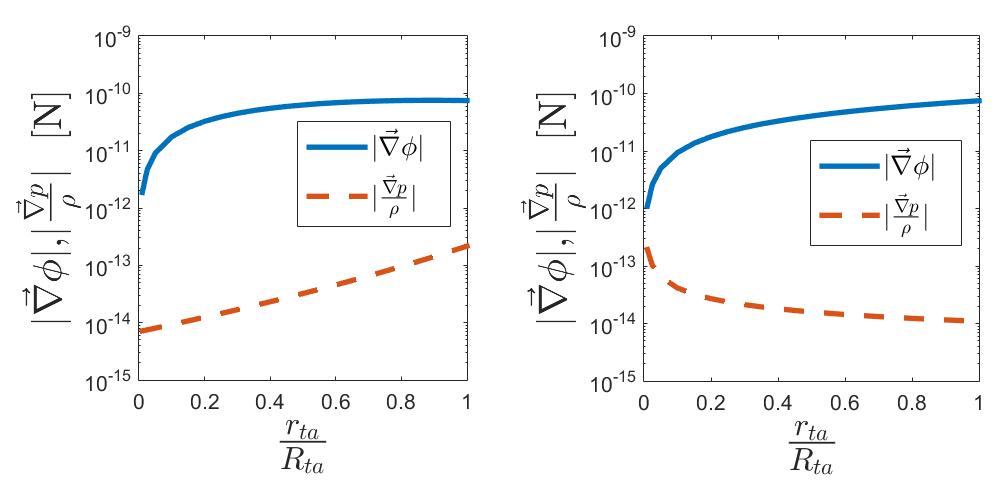}
\caption{Same as in Fig.~\ref{Beta_7_kpc.jpg}, but for 
$\beta = 1$  (left) and $\beta = 0.1$ (right). The difference with aforementioned 
figure reflect situations whereby the initial profile is linearly decreasing 
with radius ($\beta = 1)$, or approximately flat except near the center ($\beta = 0.1$). The initial density gradients are smaller than for $\beta = 7$ in the outer 
regions.} 
\label{Beta_1_0.1_kpc.jpg}
\end{figure*}

Assuming the aforementioned maximal average density contrast 
fixes the first factor on the right hand side of (\ref{eq:denprof}).
Requiring  $\frac{\bar{\delta}_i(x = 1)}{\bar{\delta}_i(R_i)} = 1$,
fixes $c$ to 
$c = 1-[\frac{\delta_i(0)}{\bar{\delta}_i(R_i)}]^{-1}$.
As noted, the 
average overdensity 
$\bar{\delta}_i(R_i)$ may be presumed to correspond 
to the linear $\sigma_M (R_i, z_i)$. 
To ensure such a correspondence --- that overdensities are adequately 
linear at all radii $r_i (z_i)$ within the sphere --- we start  
our evolution at $z_i = 300$ (larger than the starting redshift in the previous section). 
Given  $\beta$, this fixes 
the profile completely.

The magnitudes of the pressure and gravitational forces
are given by
\begin{equation} |\bm{\nabla} \phi| = \frac{G M} {r^2} 
\end{equation}
\begin{equation} 
\bigg|\frac{\bm{\nabla}P}{\rho}\bigg| = \frac{1}{\rho} \frac{dP}{d\rho} \frac{d\rho}{dr}  = \frac{Ac^2} {\rho} \frac{d\rho}{dr_{ta}}, 
\label{eq:pressuregrad}
\end{equation}
where we have assumed $\alpha = 1$ in Eq.~(\ref{eq:state}), as for a fiducial Chaplygin 
fluid configuration. 
If the pressure forces are negligible along an evolution 
described by~(\ref{eq:cycloid}),  
the mass
inside each shell is practically conserved and 
\begin{equation} 
M  = M (<r_i) = 4 \pi \int_0^{r_i} r^2 \rho_i dr,
\end{equation}
which is readily evaluated given~(\ref{eq:denprof}). The local 
density is then given by 
\begin{equation} 
\rho = \frac{1}{4 \pi r^2} \frac{dM}{dr}. 
\end{equation}
With negligible pressure (and no shell crossing), $M = M (<r) = M (< r_i)$, and the density will
depend only on the evolution of the Lagrangian coordinate 
$r$ and its first derivative with respect to the initial condition $r_i$.
Or, equivalently, on the change of volume between shells, as we
detail in Appendix\ref{app:sol},  where we derive the pressure force 
along the unperturbed (gravity dominated) solution~(\ref{eq:cycloid}).  

Our approach will be to assume the unperturbed solution (\ref{eq:cycloid}), calculate the 
pressure and gravity forces along it as described, 
and contend the scheme to be self consistent if we find 
$\bigg|\frac{\bm{\nabla}P}{\rho}\bigg|/ |\bm{\nabla} \phi| \ll 1$ for all ($r, t$), or equivalently ($r_i, \theta$), 
prior to shell crossing. In this case, the solution should hold to a good approximation, 
allowing for collapse into self gravitating halos.

\subsubsection{Conditions at turnaround and the nonlinear collapse scale}
\label{sec:results_nonlin}

As noted in Section~\ref{sec:basic_id}, the magnitudes of the 
gravitational forces driving the collapse are minimal at turnaround, while 
the competing pressure forces, hindering the collapse, are expected to be 
maximal near turnaround, as the average density inside a shell 
is minimal there. As these latter forces are determined by the local (as opposed 
to average) density and gradient, this statement can only be of approximate
validity. However the calculations of Appendix~\ref{app:sol} show that this is generally a good 
approximation. We will therefore display our results here at turnaround, relegating the 
full evolution plots to the appendix, while contending that  
$\bigg|\frac{\bm{\nabla}P}{\rho}\bigg|/ |\bm{\nabla} \phi| \ll 1$ at turnaround 
points to the possibility of self gravitating collapse and clustering 
in fiducial cosmological Chaplygin gas.

Figure~\ref{Beta_7_kpc.jpg} shows the results when the initial conditions 
correspond to the profile in (\ref{eq:denprof}) with $\beta =7$, with 
an overdensity with boundary $R_i = 1~{\rm kpc}$ (comoving). 
As is clear, the gravitational force is dominant at turnaround 
for all shells (at least by about an order of magnitude). We thus conclude that 
self-gravitating collapse is possible in this case. 
Fig.~\ref{Beta_1_0.1_kpc.jpg} shows 
the situation when $\beta = 1$ or $\beta =0.1$. These values correspond to profiles with 
initial density gradients that are  less steep, when $x \gtrsim 0.5$, than in 
the former case; 
for $\beta = 1$ the density falls linearly with radius, 
while for $\beta = 0.1$ it is nearly flat except in the central region. 
As can be seen from the figure,  the gravitational force is largely unaffected by the change in profile. 
But the pressure forces are smaller 
in regions where the initial profiles has larger density or smaller gradient.  
And, again,  these forces are invariably far smaller in magnitude
than the gravitational ones, thus allowing for 
self gravitating collapse.

\begin{figure*}[!htbp]                                     
\centering
\includegraphics[width=1\textwidth]{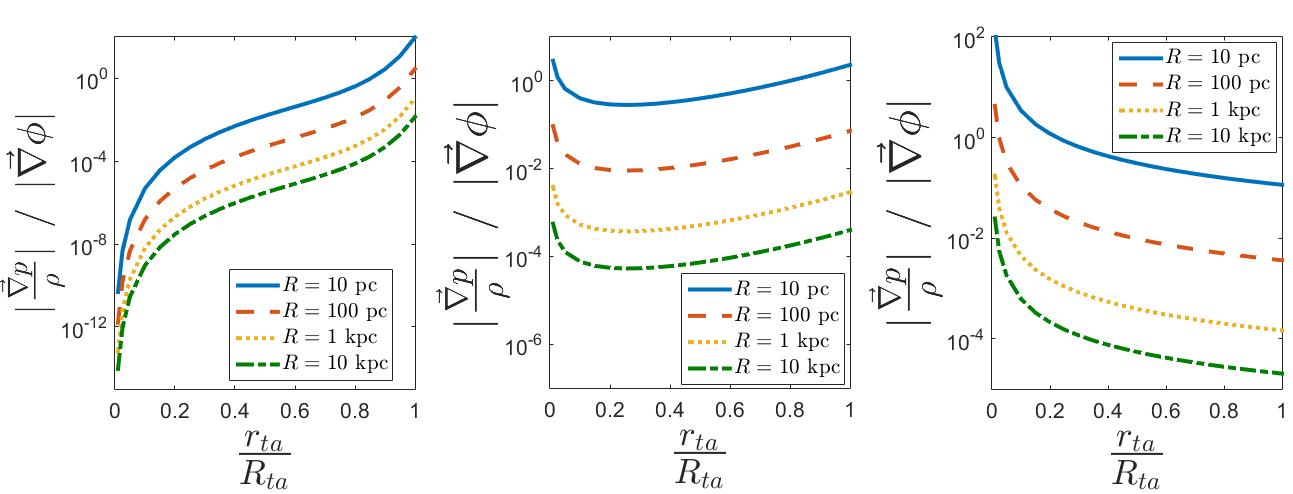}
\caption{The ratio between of the pressure to gravitational forces 
at turnaround for the profiles in~(\ref{eq:denprof}), with 
$\beta$ = 7, 1 and 0.1 (from left to right). The perturbations 
are probed at the various scales indicated.  The transition scale 
at which the ratios switch from values smaller to ones larger than unity 
(at $R \approx {1 \rm kpc}$)
constitutes what we have termed the 'nonlinear Jeans scale'. 
}
\label{Ratio.jpg}
\end{figure*}

Fig.~\ref{Ratio.jpg} shows the ratio of the magnitudes of the 
pressure to gravitational forces when the initial perturbation
radii $R_i$ correspond to the various indicated spatial scales. 
It shows that, at smaller scales, the pressure forces can become large enough to 
impede collapse, but that they gradually diminish relative to the 
gravitational force as the scale increases. The transition to a regime 
where they may be considered negligible 
generally occurs at an initial (comoving) perturbation scale of order ${\rm 1~kpc}$. 
Thus, while the results of Section~\ref{sec:mar_stab} (particularly Fig.~\ref{fig:SigmaM_A1}),
show that all the scales should be Jeans stable 
in the linear regime, one may now in define a "nonlinear Jeans scale" at the 
kpc scale. This phenomenon, absent in regular gases, arises from the 
peculiar property connected to the gas at hand, namely the decreasing magnitude 
of the pressure with density. 

From these results, it is clear that  the collapse of comoving smoothing scales of $R \approx 100~{\rm kpc}$, 
corresponding to halo masses of order $10^8 M_\odot$, 
may be  readily realized as in CDM. This is likely to be the smallest halo mass scale relevant to observed dwarf satellite galaxies~\cite{Nadler20}. On the other hand, interestingly, 
significantly smaller scales would be progressively suppressed, which  may be of relevance to the small scale problems associated with CDM~\cite{delpopolo, bullock}.

\section{Backreaction on background dynamics}
\label{sec:bback}

\subsection{General considerations}

The results of the previous section show that it is quite plausible 
for nonlinear clustering to occur in the standard Chaplygin gas.
We here examine the effect of this phenomenon on the background evolution. 
As  noted in~\cite{2014PhRvD..89j3004A}, this may bring that evolution 
closer to that of the 
largely successful standard $\Lambda$CDM scenario. 
We illustrate this here in simplest terms, through  modification 
of Eq.~(\ref{eq:density}). For concreteness, we focus on the standard ($\alpha =1$) case, 
though the results may be trivially generalized 
We thus rewrite Eq.~(\ref{eq:density}) as 
\begin{equation}
\rho_{\rm Ch} (a) = \left(A_{\rm cl} + \frac{B_{\rm cl}}{a^6} \right)^{1/2}.     
\label{eq:rhoclus}
\end{equation}
As with $A$ and $B$, the new coefficients $A_{\rm cl}$ and $B_{\rm cl}$ must be 
calibrated to be compatible with observations. At recombination ($a \sim 10^{-3}$), the first term in brackets 
is negligible (suppressed by a factor $a^6$ compared to the first).

In order to fit CMB data, with this component representing a matter contribution~\footnote{For simplicity, we ignore 
here the baryon contribution, which will be included in the comparison with observation in the following subsection}, 
one should have
\begin{equation}
  B_{\rm cl} = B = \rho_m^2 (a = 1) = \rho_c^2 \Omega_m^2. 
\end{equation}
Then suppose that at $a = a_{\rm cl} <  10^{-1}$ 
the smallest scale that can cluster, in accordance with calculations of the previous section, collapses into halos. A  phase transition then occurs, with the clustered phase behaving thereafter as a pressureless component made of collapsed halos, which may henceforth hierarchically merge. We assume that a definite fraction of Chaplygin gas $f$ splits into this CDM-like component. The density of the unclustered component then decreases by a factor $1-f$, but unless the clustering is so efficient, such that $1-f < 10^{-3}$, the first term in the bracket of~(\ref{eq:rhoclus}) remains negligible.  
The density of the system then splits into
\begin{equation}
\rho  = \rho_{\rm Ch} + \rho_{\rm CDM} = (1-f) \frac{B^{1/2}}{a_{\rm cl}^3} + f  \frac{B^{1/2}}{a_{\rm cl}^3}. 
\end{equation}
For $f > 0.5$, most of the energy density is in 
the clustering component (with density $\rho_{\rm CDM}$) at this point, while a minority remains 
in the form of homogeneous Chaplygin gas (with $\rho_{\rm Ch}$).  
But at late times (as $a \rightarrow 1$), the first term in Eq.~(\ref{eq:rhoclus}) becomes important,  
and the total density (of homogeneous Chaplygin plus clustered CDM-like component) evolves as
\begin{equation} 
\rho   =  \left(A_{\rm cl} + \frac{(1-f)^2 B}{a^6}\right)^{1/2} + f   \frac{B^{1/2}}{a^3}.    
\label{eq:rhoclusfull}
\end{equation}

\begin{figure}[]
\includegraphics[width=0.48 \textwidth]{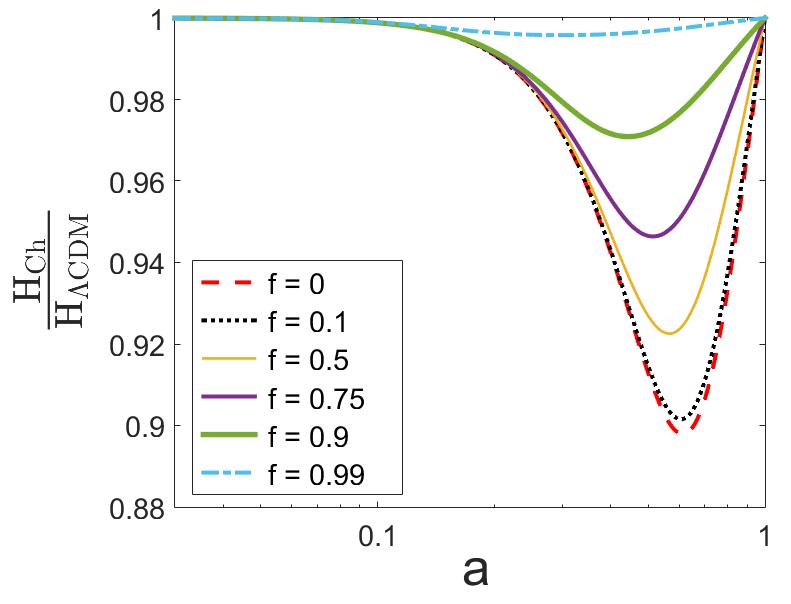}
\caption{Ratio of the evolution of the Hubble parameter
in clustered standard  Chaplygin gas universes ($\alpha =1$ in Eq.~\ref{eq:state}), to that 
of $\Lambda$CDM, with $\Omega_m = 0.3$ and $\Omega_\Lambda = 0.7$, 
for various values of the fraction $f$ of the clustered (CDM-like) 
Chaplygin gas component.}
\label{fig:back}
\end{figure}

Evaluating this at $a=1$, and assuming a current critical density $\rho_c$, 
gives
\begin{equation}
    A_{\rm cl} = A + 2 f \rho_c^2 \Omega_m^2 \left(1- \frac{1}{\Omega_m}\right),
\end{equation}
where $A= (1- \Omega_m^2) \rho_c^2$ (from Eq.~\ref{eq:ABH}).

For $f \rightarrow 0$, $A_{\rm cl} \rightarrow A$, as expected.  
It is worthwhile to note, however, that for $\Omega_m \approx 0.3$ 
and a high level of clustering,  $A_{\rm cl}$ is 
significantly smaller than $A$. 
As a Chaplygin dark sector that can support
the observed large scale structure must embody a clustered medium. 
It would  therefore be characterized by $A_{\rm cl}$, rather than $A$, in its equation of state. 
The calculations of the previous sections, where $A$ was used as customary, thus involve larger sound speeds and pressure gradients. They should actually be considered  conservative in estimating the relative strengths of the gravitational 
to pressure forces.

Because the terms suppressed by powers of $a$ rapidly decrease with $a$, there 
are two transitions embodied in the form of Eq.~(\ref{eq:rhoclusfull}). One occurs when 
$A_{\rm cl} \gtrsim \frac{B}{a^6} (1-f)^2$. This indicates that the non-clustered 
Chaplygin component has transited to a dark energy like fluid, behaving like 
a cosmological constant, with density $\rho_\Lambda = \sqrt{A_{\rm cl}}$. 
With $\Omega_m =0.3$ and $f \rightarrow 1$, this nicely gives
$\rho_\Lambda \rightarrow 0.7 \rho_c$. 
This is not surprising, as the 
advent of cosmological constant-like component
renders the background evolution 
closer to  $\Lambda$CDM than that of the homogeneous Chaplygin gas cosmology.
We illustrate this in Fig.~\ref{fig:back}, where we plot the ratios 
of the Hubble parameter in flat clustered Chaplygin gas universes
to those of $\Lambda$CDM, assuming the same 
values $H_0$ for all models at $z=0$, 

The transformation of the homogeneous component into $\Lambda$-like
sector occurs when
\begin{equation}
a_\Lambda \gtrsim \left(\frac{B}{A_{\rm cl}}  \right)^{1/6} (1-f)^{1/3}, 
\end{equation}
which happens at earlier $a$ (by a factor $\sim (1-f)^{1/3}$),
compared to the homogeneous Chaplygin gas scenario. The latter 
is still in transition at $a =1$, causing tension with late universe background cosmic dynamics~\cite{2014PhRvD..89j3004A}. 

Finally, we have the transition whereby the total cosmic energy density 
switches from being dark matter dominated, to dark energy domination. 
By this time the homogeneous Chaplygin component has already fully transited into 
its $\Lambda$-like phase. If we accordingly ignore the second term in 
Eq.~(\ref{eq:rhoclusfull}), a clustered Chaplygin gas universe should 
transit to a dark energy dominated regime at 
$a \gtrsim f^{1/3} \left(\frac{B}{A_{\rm cl}}\right)^{1/6}$. 
As the equation of state of the homogeneous dark energy component 
is now effectively that of a cosmological constant with $p = -\rho_\Lambda$, 
imposing the condition $\rho + 3 p =0$, 
leads to a deceleration-acceleration transition at
\begin{equation}
    a_{\rm da} \gtrsim \left(\frac{f}{2}\right)^{1/3} \left(\frac{B}{A_{\rm cl}}\right)^{1/6}.
\end{equation}
With $\Omega_m =0.3$ and $f \rightarrow 1$, this tends to 
$a_{\rm da} = 0.6$, with $z_{\rm da} = 0.67$, which is compatible  with
lower bounds inferred from observations. 
The transition redshift increases as $f$ decreases.
It  remains within viable bounds $z_{\rm da} \lesssim1$
as long as the clustering is efficient ($f \gtrsim 0.7$). 

One can thus count three transitions. The first one  leads to a clustered medium, whereby part of the Chaplygin gas collapses into halos and acts henceforth as a CDM-like component; the second transition occurs when the remaining homogeneous gas starts to effectively act as a cosmological constant; while the third transition takes place when the 
Chaplygin universe transits to an accelerated phase. This latter development occurs
much the same way as $\Lambda$ becomes dominant in the standard model. Furthermore, when the 
matter energy density is calibrated to the CMB, 
the $\Lambda$-like component comes with energy density 
compatible with its measured late time value in the context of the $\Lambda$CDM model.
The basic characteristics of the clustered Chaplygin cosmology 
appear in this context quite akin to $\Lambda$CDM, including a similar transition time for its transformation from decelerated to accelerated expansion.

\subsection{Constraints from observations}
\label{sec:obs}
\begin{figure*}[]
\centering
\includegraphics[width=1. \textwidth]{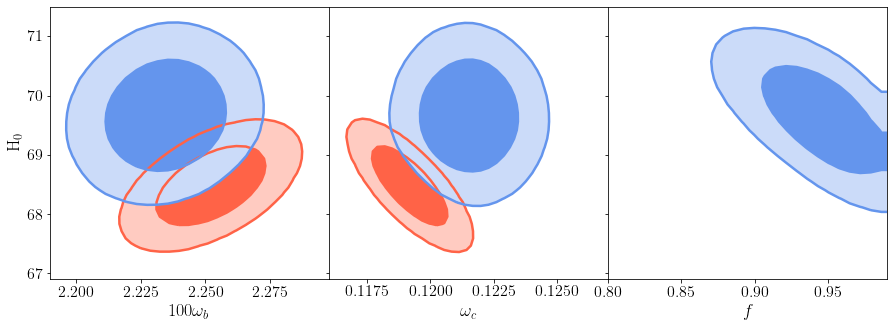}
\caption{One and two sigma contours of the Hubble parameter and physical 
baryon and dark matter densities at $z=0$, in flat clustered Chaplygin 
gas universes (blue) and $\Lambda$CDM (red). In the former case we also 
show the distribution of the clustering fraction $f$ against $H_0$.  
The dataset used includes  SNe~Ia, CMB and BAO data, as well as 
cosmic chronography (CC) estimates of $H(z)$. The  
$\chi^2$ for the best fitting $\Lambda$CDM and Chaplygin models are 1054 and 1048, 
respectively (with corresponding reduced $\chi^2/{\rm dof}$ of 1.027 and 1.021.}
\label{fig:constraints}
\end{figure*}

In order to obtain  quantitative constraints on $f$, we
confront the model with observational datasets constraining the background cosmology.  
For this purpose we use supernova SNe~1a, baryon acoustic oscillation (BAO) and cosmic microwave background (CMB) data, in addition to cosmic chronography (CC) estimates of the Hubble parameter $H (z)$.

For supernova data we use the Pantheon sample~\cite{Pan-STARRS1:2017, Pantheon21}, 
combining data from the Pan-STARRS1 Medium Deep Survey 
with older observations, for a total 1048 SNe~Ia  in the redshift range $0.01< z < 2.3$. 
We also use anisotropic BAO data from the Sloan Digital Sky Survey (SDSS-III), as provided in~\cite{Alametal17},
to simultaneously constrain the Hubble parameter and the angular diameter distance. 
Additional constraints on $H(z)$ 
come from CC data obtained from~\cite{CCMagana:2017nfs}.  
As the CMB spectrum at last scattering ($z = z_*$) is 
practically unaffected by the modification of late time dark energy 
behavior introduced by replacing $\Lambda$CDM by a Chaplygin gas dark sector 
(as the contribution to the total energy density from the dark energy-like  
component at $z_*$ is of order $1/z_*^3 \lesssim 10^{-9}$ in both cases),   
constraints on the background evolution 
may be effectively expressed by a 'compressed likelihood' of the CMB
power spectrum~\cite{Planck2015DE, ShiftpriorZhai:2019nad}.  This includes the shift parameter 
$\mathcal{R} = \frac{1}{c} D_M (z_*) \sqrt{\Omega_m H_0^2}$, where
$D_M$ is the comoving angular diameter distance to last scattering, which in a flat FRW universe is given by
\begin{equation}
    D_M (z_*) = c \int_{0}^{z_*} \frac{d z}{H (z)}. 
\label{eq:codiam}
\end{equation}
Another CMB distance prior is 
the angular scale of the sound horizon 
at last scattering $l_A = \pi D_M (z_*)/r_s(z_*)$, where $r_s (z_*)$ is the comoving sound 
horizon at $z_*$. Finally, an additional constraint 
is obtained from the present physical baryon density $\omega_b = \Omega_b \rho_c$.  
We use the following values, based 
on an analysis of the Planck-2018 TT, TE, EE+lowE data~~\cite{Chenetal2018}: 
$\mathcal{R} = 1.7502 \pm  0.0046$,
$l_A = 301.471^{+0.089}_{-0.090}$ 
and $\Omega_b h^2 = 0.02236 \pm  0.00015$. 
We note the particularly strong constraint on
$l_A$;~$|\Delta l_A| /l_A \simeq 3 \times 10^{-4}$.

To enable proper comparison with those datasets, the equations of the 
previous subsection are modified, so as to account for a separate baryonic 
and radiation components. Thus $B$  now refers 
just to the dark matter density, so that $B_{\rm cl} = B = \rho_c^2 \Omega_c^2 = \omega_c^2$,  and Eq.~(\ref{eq:rhoclusfull}) 
has two additional additive terms, incorporating the baryon and 
radiation energy densities; namely $\Omega_b/a^3$ and $\Omega_R/a^4$. 
Also, by defining $\rho (a=1) = \rho_c$, one now has 
\begin{equation}
A_{\rm cl} = \rho_c^2 \left[\Omega_{\rm Ch}^2  - (1-f)^2 \Omega_c^2 \right],
\end{equation}
where $\Omega_{\rm Ch} = 1- (f \Omega_c + \Omega_b + \Omega_R)$.

The modified equation~(\ref{eq:rhoclusfull}) is then 
included into the publicly available code CLASS~\cite{CLASS2011arXiv1104.2932L}, 
in order to evaluate the theoretical predictions.  
The statistical analysis 
is conducted by  running a  Monte Carlo Markov chain over models, using 
the publicly available emcee MCMC code~\cite{MCMC2013PASP..125..306F}.

Keeping the number of relativistic degrees of freedom fixed as in 
standard cosmology, the   
background evolution of the clustered Chaplygin 
gas universe is determined by four parameters. 
Three of these --- $H_0$, $\omega_b$, $\omega_c$ --- are shared with $\Lambda$CDM, with which we will be comparing our clustered Chaplygin gas models, by conducting control MCMC runs. 
For these parameters, we take flat priors in the following forms: 
$H_0 = [65: 80]~{\rm km/s/Mpc}$,  $\Omega_c h^2 = [0.1:0.131]$ and $\Omega_b h^2
= [0.015:0.028]$. 
A prior is also placed on the absolute magnitudes of the SNe~Ia: $M= [-20: -18]$.  

The additional parameter in our 
clustered Chaplygin model is the clustering fraction $f$. When $f \rightarrow 0$, one recovers the conventional Chaplygin gas, which has been shown to be strongly disfavored 
by observations (recall that we are assuming, throughout this section, that we are dealing with 
a standard Chaplygin gas with $\alpha =1$ in Eq. ~\ref{eq:state}). 
Here we want to examine the likelihood of clustered models characterized 
by different $f$. For this purpose, we place a flat prior 
encompassing the whole range of possibilities $f = [0:1]$.  
The posterior probabilities of 
all parameters are then obtained from the MCMC analysis.

\begin{figure}[]
\includegraphics[width=0.48 \textwidth]{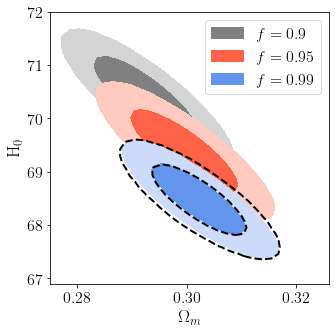}
\caption{Convergence of clustered Chaplygin models towards $\Lambda$CDM
as the clustered fraction $f \rightarrow 1$. Dashed contours refer to 
$\Lambda$CDM. Note the systematically  larger 
$H_0$ for smaller $f$.}
\label{fig:Convergence}
\end{figure}

The results are shown in Fig.~\ref{fig:constraints}.  
The best fit values for the clustered Chaplygin gas is $f \simeq 0.95$. 
Values down to $f = 0.91$ are allowed at the 1-sigma level. 
Slightly smaller values of $f \gtrsim 0.88$ are allowed at  2-sigma.
The background evolution of the clustered Chaplygin gas may thus be considered viable for these clustering levels and associated levels of confidence. 
Indeed, we find the total $\chi^2$ is slightly smaller for the best fitting 
clustered Chaplygin gas than the best fitting $\Lambda$CDM
(1054 versus 1048). 
On the other hand, the
unclustered standard Chaplygin gas model, with $f \rightarrow 0$
(and $\alpha =1$), 
is clearly ruled out, as expected. At the opposite limit, 
as $f \rightarrow 1$, the model tends towards $\Lambda$CDM, 
as expected from the discussion of the previous subsection, and 
illustrated in Fig.~\ref{fig:Convergence}.

\begin{figure}[]
\includegraphics[width=0.48 \textwidth]{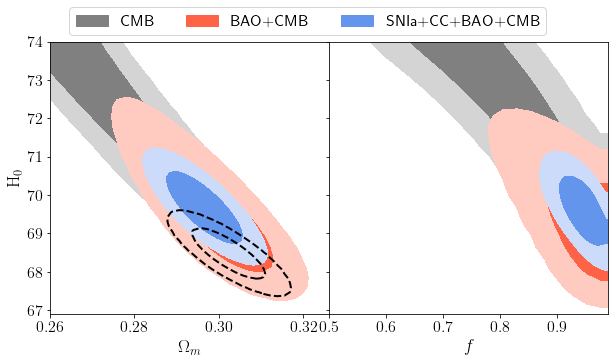}
\caption{Constraints from the various datasets.
While $\Lambda$CDM is highly constrained by CMB data, 
clustered Chaplygin gas models are compatible 
with larger $H_0$ if the clustered fraction $f$ is relatively small.  
Smaller values of $f$ (and larger $H_0$) are however disfavored 
by other datasets (primarily by high precision 
intermediate distance measurements coming from the 
BAO). Thus favored models remain close to $\Lambda$CDM (represented by 
the dashed contours).}
\label{fig:Combos}
\end{figure}

\begin{figure}[]
\includegraphics[width=0.48 \textwidth]{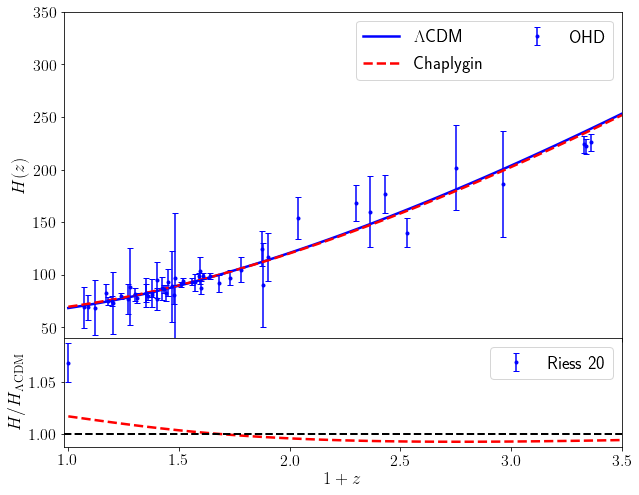}
\caption{Observational Hubble data (OHD), from CC and BAO measurments,  
and best fitting $\Lambda$CDM and 
clustered Chaplygin models. The local value of $H_0$, as measured
by Riess et. al.~\cite{Riessetal:2020fzl}, is also shown in the lower panel
(but not included in the likelihood analysis of either model).}  
\label{fig:Riess}
\end{figure}

The mean value of $H_0$ for the clustered Chaplygin 
models is larger than that obtained for $\Lambda$CDM:  
namely $69.7 {\rm km/s/Mpc}$ for the former versus 
$68.5 {\rm km/s/Mpc}$ for the latter, when $f$ is left free.   
When $f$ is fixed, $H_0$
is also systematically higher for smaller $f$ (Fig.\ref{fig:Convergence}). 

The larger $H_0$ in the Chaplygin case can be understood in terms of the behavior of the Hubble parameter shown in Fig.~\ref{fig:back}, and noting again that the acoustic scale $l_A$ is the most strongly constrained observable in our analysis. 
This principally translates into a constraint on the  comoving  
angular diameter distance to the CMB last scattering surface. 
In $\Lambda$CDM this distance is entirely fixed. given 
the physical energy densities associated with radiation, dark matter, 
baryons, and the cosmological constant.  
But since (from Fig.~\ref{fig:back}),  
when $H_{\rm Ch}/H_{\rm \Lambda CDM}$ is set to unity at $z=0$,  
for $z  > 0$ 
and $f < 1$ there are 
intervals for which the ratio is smaller than unity, it follows from Eq.~\ref{eq:codiam} that one can obtain the same value for $D_M$ by 
rescaling  $H_{\rm Ch} (z)$, such that 
$H _{\rm Ch} (z=0)/ H_{\Lambda {\rm CDM}} (0) > 1$.

This situation is similar to the case of phantom-like dark energy models.
And, as in that case, one may actually entirely
`solve' the $H_0$ tension, if this is solely 
defined as a tension between
local and CMB measurements of $H_0$. 
However, also as in the phantom case, 
distance measures on the way to the CMB, particularly high precision 
BAO data, constrain the models to be close to $\Lambda$CDM~\cite{El-ZantHanaf:2018}.  
In the present context
this requires $f$ to be close to unity, and thus $H_0$ to be only 
slightly larger than in $\Lambda$CDM, compared to what is needed 
to entirely alleviate the $H_0$ tension. 
The behavior of the ratio of $H (z)$ for the best fitting 
Chaplygin and $\Lambda$CDM models, reflecting the competing constraints, 
is shown in the lower panel of~Fig.~\ref{fig:Riess}.

\section{Conclusion}
\label{sec:level4}

Unified dark matter models are appealing for combining 
the dark sector in a single component, which can act as both 
dark energy and dark matter. The prototypical example is that 
of the generalized Chaplygin gas, with equation of state~(\ref{eq:state}), 
providing for small pressure at high densities
and significant negative pressure at low density. However, 
unless the generalized Chaplygin gas parameter is chosen such that the associated 
cosmology 
is virtually indistinguishable from $\Lambda$CDM, or superluminal sound speeds
are allowed, linear perturbations in a homogeneous Chaplygin fluid 
will become Jeans stable and oscillate acoustically and damp, rather than grow, on 
scales observed in large scale structure surveys. This, in particular,
is true for the theoretically  motivated standard Chaplygin gas with 
$\alpha =1$. 

Here we first note that, in a hierarchical structure formation model, 
dark matter perturbations probed by large scale structure surveys do 
not occur in a homogeneous fluid. Rather, they occur in a medium 
that is already hierarchically clustered, starting from dwarf galaxy scales or smaller). 
It is sufficient for the Chaplygin fluid 
to collapse into halos early on, on some small scale (dwarf galaxy halo or below), 
for a CDM-like component to materialize and cluster hierarchically. 

We thus ask whether any small scale seed perturbations can grow sufficiently 
to collapse early on. 
We first examine the linear stability and 
check whether perturbations may grow to reach 
the usual Press-Schecter threshold required for nonlinear collapse. 
We find that while this may be the case 
for very small or large values of $\alpha$, such an analysis suggests 
that acoustic oscillations generally prevent the achievement of the threshold. 
However, for the fiducial case of $\alpha =1$, the RMS 
fluctuations do come quite close to achieving 
the critical threshold for scales associated with dwarf galaxies and below.

That the perturbations come tantalizingly close to crossing the critical 
collapse threshold for this relatively well motivated model --- and at 
the minimal scales expected in a successful hierarchical collapse model, with no free parameters tuned --- suggests that a nonlinear analysis is warranted. This is 
particularly motivated by the fact that pressure forces in a unified dark matter 
fluid decrease in magnitude with increasing density (characteristic of nonlinear phase), 
as opposed to the case of laboratory fluids where the pressure and its gradients 
become more important as the density increases. Thus systems that are stable against self 
gravitating collapse in the linear regime may not be so in a nonlinear analysis. 

Numerical and theoretical modelling of the dynamics 
of a negative pressure fluid with sound speed decreasing with density
is largely unexplored. Through a simple secondary infall model, we
attempt to circumvent expected difficulties, 
while capturing the basic ingredients that determine the possibility 
of collapse and formation of self gravitating objects in the nonlinear regime. 
We choose the initial density distribution to correspond to a system 
where all shells turnaround, before any shell crossing occurs, when the 
system is evolved solely under gravity. 
Then we evolve the dynamics, starting in the linear regime at high redshift. 
We show that while a linear analysis  predicted that kpc scale perturbations 
were marginally (Jeans) stable, nonlinear evolution suggests 
that pressure forces should in fact be negligible compared to gravitational ones 
along the whole inhomogeneous tophat trajectory, for perturbations on scales
$\gtrsim 1 {\rm kpc}$, thus enabling gravitating collapse.

One may, in this context, define a 'nonlinear Jeans scale', which does not 
have a counterpart in standard gases. It arises from the peculiar characteristic 
of the decreasing magnitude of the pressure in Chaplygin gases. 
Nonlinear perturbations larger than this Jeans length can also grow, 
as the ratio of the gravitational forces becomes smaller still. 
In particular,  the collapse of comoving  scales 
associated with small dwarf satellite galaxies, should be readily allowed.

Once structures form on  small scales, a CDM-like component is present. 
It may proceed to cluster hierarchically, independently of the remaining homogeneous 
Chaplygin fluid. The latter may still act as dark energy. The acoustic 
oscillations,  thought to rule out unified dark  fluid models, would only be imprinted
in that dark energy component and would not appear in the galaxy power spectrum, which 
would correspond to CDM.

The formation of structure on scales smaller than the nonlinear Jeans scale, on the other hand,  would be suppressed.   
This may be of relevance to the small scale problems associated with CDM; for example,
the apparent overabundance of small CDM halos compared to the number of 
small galaxies observed. As with particle CDM 
alternatives, devised in part to address such problems, a Chaplygin gas model may be tested 
with observations of small scale structure~\cite{Nadler21PhRvL, Nadler21}.

In light of these results,  the problem of acoustic oscillations in the linear power spectrum 
of Chaplygin gases may not be as serious as usually assumed, provided  the hierarchical 
structure formation process is adequately taken into account. 
In particular, it would appear less serious than the problem of finding self gravitating equilibrium with a density 
distribution corresponding to that inferred from observations at large radii around galaxies and clusters, while keeping  the same equation of state\cite{El-ZantChap15}. 

Furthermore, in the context of the present analysis, the 
basic characteristic background evolution of the clustered Chaplygin gas cosmology is found to tend towards $\Lambda$CDM  as the clustering efficiency is increased: 
when clustering occurs, the remaining  homogeneous component constitutes a 
cosmological constant-like sector early on, with  an energy density 
akin to the corresponding one in $\Lambda$CDM; the acceleration deceleration 
transition occurs also as in the standard models. Finally, the pressure forces associated with the clustered 
gas are  smaller (through a rescaling of the parameter $A$ in Eq.~\ref{eq:state}, as described in Section~\ref{sec:bback}). 
This reinforces the consistency of the collapse model, and 
may also help alleviate the aforementioned problems related to the dynamics of the outer parts of galaxies and clusters. 

Quantitative comparison with 
observational datasets shows the  background dynamics of the 
clustered Chaplygin gas models to be viable (at the one sigma level)
if the fraction of fluid that collapses into a 
CDM-like component of small halos (that may subsequently hierarchically cluster) is larger 
than $90 \%$. 
The associated value of the Hubble constant is larger than in $\Lambda$CDM, 
due to effects similar to those present in 
phantom dark energy models. As in these models,  the 'Hubble tension' 
may be completely resolved if it is defined solely in terms of discrepancy 
between local and CMB measurements. But other data, particularly high accuracy 
BAO distance measurements, dictate that viable models 
(and their $H_0$) remain relatively close to $\Lambda$CDM. 

 Further investigation, beyond the simple nonlinear collapse model presented here,  
 requires an examination of what happens at shell crossing in the later stages of self gravitating collapse, 
 including the treatment of possible shocks. 
 Though some work regarding the Riemann problem for Chaplygin gas exists in the 
 mathematical literature, including idealized shock simulations~\cite{2009ArRMA.191..539S, Wang2013TheRP, YANG20125951}, 
 the physical consequences of the phenomenon 
 remains unexplored. To our knowledge, no
 detailed numerical simulations of the dynamics of any Chaplygin fluids 
 have been conducted, much less of its self gravitating cosmological evolution.  
 We hope that the proof of principle presented here, suggesting that the problem of large scale oscillations in the power spectrum
 should not be as insurmountable
 as widely believed, would help reopen detailed investigation of the 
 consequences of structure formation in the context of Chaplygin gas cosmologies.

\begin{acknowledgments}
We would like to thank Waleed El Hanafy and especially 
Mahmoud Hashim for discussions and help with Section~\ref{sec:obs}. 
This project was supported financially
by the Science and Technology Development Fund (STDF),
Egypt. Grant No. 25859 and Grant No. 33495. 
\end{acknowledgments}

\appendix
\section{Ratio of pressure to gravity force along parametric solution}
\label{app:sol}

\begin{figure*}[!htbp]
\centering
\includegraphics[width=1\textwidth]{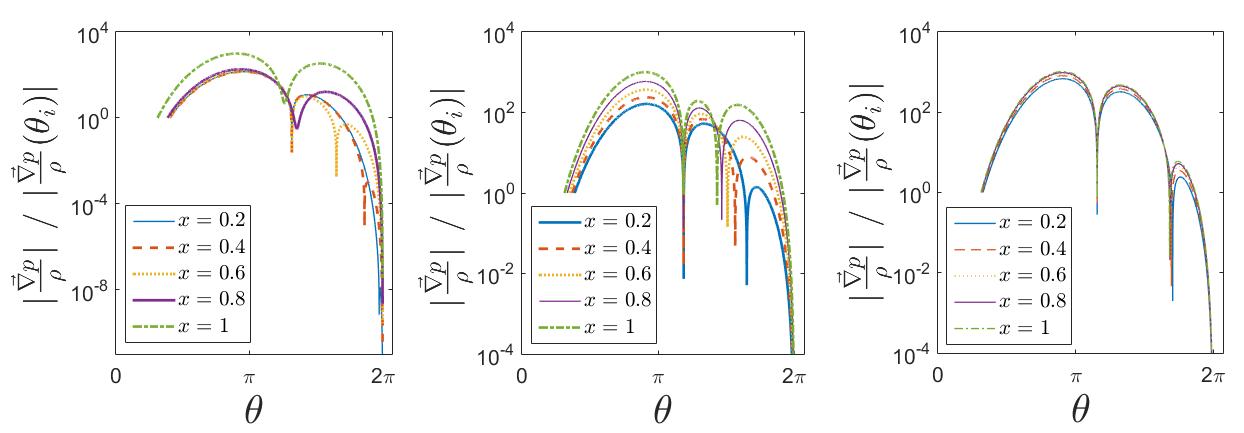}
\caption{Scaled pressure along parametric 
solution~(\ref{eq:cycloid})  at various values of the 
scaled initial radius 
$x = r_i/R_i$. The initial density profiles are given by (\ref{eq:denprof}) with (from left to right)  
$\beta$ = 7, 1 and 0.1 and $R_i = {\rm 1~kpc}$, starting at 
$z_i= 300$ with overdensity equal to the RMS fluctuation of the linear Gaussian field: 
$\bar{\delta} (R_i) = \sigma_M (R_i, z_i)$. 
The derivative discontinuities (reflecting sign switching in $d^2r/dr_i^2$ and $dr/dr_i$),
correspond to consecutive shells approaching instead of increasing their separation, 
and then to eventual shell crossing. 
The calculation of the pressure beyond this (shell crossing)
point is only formal, as the dynamics reflected in the solution (\ref{eq:cycloid}) no longer 
strictly apply. Note that the steeper the change in density through the bulk of the system 
(larger $\beta$), the more spaced out the discontinuity angles for different values of $x$.}
\label{Pressure_Scaled_2.jpg}
\end{figure*}

\begin{figure*}[!htbp]
\centering
\includegraphics[width=1\textwidth]{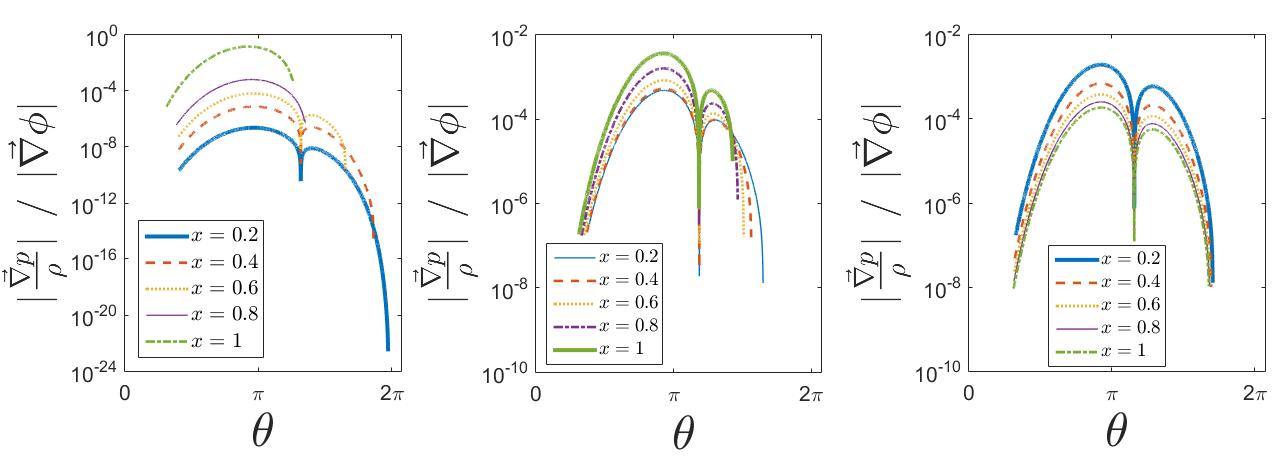}
\caption{Same as in Fig.~\ref{Pressure_Scaled_2.jpg}, but
for the ratio of  pressure to gravitational force, and  
until local shell crossing occurs.}
\label{Ratio_2.jpg}
\end{figure*}

We wish to evaluate the relative magnitude of the gravity to pressure forces 
along the  cycloid solution (\ref{eq:cycloid}). 
As this solution reflects evolution solely under the 
influence of gravity (and with no shell crossing), 
it would still approximately hold 
if the pressure forces remain small relative to gravity along it, at all 
temporal stages and spatial radii.  

Before shell crossing occurs, finding the gravitational force on
a shell at radius $r$ is trivial; it is simply given by $- G M/r^2$, with the mass $M$ enclosed in $r$ taken as constant, and the radius given by the first by the cycloid solution~(\ref{eq:cycloid}). 
To evaluate the pressure forces 
along this 'unperturbed' solution, we map the 
evolution of the local density and its gradient
along it.  For this purpose we consider neighboring initial 
conditions, starting at same initial time $t = t_i$
and developing according to equations~(\ref{eq:cycloid}).

In the absence of shell crossing, the evolving density $\rho (r, t)$
is given in terms of the initial distribution $\rho_i$
through the change of the volume element between neighboring shells. Namely 
\begin{equation}       
\rho(r,t) = \left \lvert \frac{dV_i}{dV} \right \rvert \rho_i (r_i,t_i),  
\label{eq:simdens}
\end{equation}
where (using the first of equations \ref{eq:cycloid}), 
\begin{equation}
\left \lvert \frac{dV_i}{dV} \right \rvert  
=   \frac{r_i^2}{r^2}   \left \lvert \frac{dr_i}{dr} \right \rvert
= \left(\frac{1 - \cos\:\theta_i}{1 - \cos\:\theta}  \right)^2
 \left \lvert \frac{dr_i}{dr} \right \rvert.
\label{eq:voldens}
\end{equation}
The density gradient can also be written as
\begin{equation}
 \frac{d\rho(r,t)}{dr} =   \frac{dr_i}{dr}  \left (\left \lvert \frac{dV_i}{dV} \right \rvert \:  \frac{d\rho_i(r_i,t_i)}{dr_i}
+ \rho_i(r_i,t_i) \: \frac{d}{dr_i} \left \lvert \frac{dV_i}{dV} \right \rvert \right).
\label{eq:densgrad}
\end{equation}

The explicit forms of the terms required to calculate the density and its gradient 
for the profiles adopted in this study are given in the following appendix.
We here present the results.
As in main text, we use the code CAMB to produce $\sigma_M (R_i)$ at $z_i = 300$, with 
$H_0 = 69~\rm{km s^{-1}  Mpc^{-1}}$. 

Fig~\ref{Pressure_Scaled_2.jpg} shows results for the scaled pressure forces on various shells 
are shown for the three chosen values of $\beta$, reflecting flatter initial outer profiles (i.e. as
$x = r_i/R_i \rightarrow 1$), for the standard Chaplygin gas ($\alpha = 1$).
As may be expected (and already noted in Section~\ref{sec:model}), given the equation of state (\ref{eq:state}), the pressure 
force increases as a shell expands, reaching a maximum near turnaround. 
It then rapidly decreases as the shell contracts. 
The derivative discontinuities correspond to switching in signs 
in the second derivatives of the radial coordinate with respect to the initial conditions
--- signaling that shells are approaching rather than increasing their separation --- 
and then in the first derivatives, when shell crossing eventually occurs.  
This occurs with wider spacing for steeper initial density gradients.  For, in the limit
of homogeneous monolithic collapse shell crossing occurs for all 
shells at same $\theta$ (and $\theta_i$ is also the same for all shells, 
as opposed to the case here).  

Beyond shell crossing our  model no longer strictly applies. 
The model also becomes inconsistent if the pressure forces ---
at any $\theta  \le \pi$, for any shell --- become 
comparable to the gravitational forces; as, in this case, 
the gravitationally dominated dynamics of the unperturbed trajectories are no 
longer a good approximation.  
As Fig~\ref{Ratio_2.jpg} shows, however, this is not the 
case. Although before turnaround ($\theta = \pi$),
the pressure force systematically increases 
and the gravity decreases, as the system expands, the ratio remains 
much smaller than unity, for all shells and all models. 
The maximum  value  of the ratio is reached 
at turnaround.  Beyond turnaround, the increasing density (hence generally decreasing pressure) 
and increasing gravity ensures that the pressure forces become smaller still relative to the gravitational ones. 
This renders self gravitating collapse possible. 
As discussed in Section~\ref{sec:results_nonlin}  (particularly in relation to Fig.~\ref{Ratio.jpg}) this conclusion is 
strengthened for larger scale nonlinear perturbations.  
The collapse of  comoving smoothing scales corresponding to a halo mass scale
relevant to dwarf galaxies. should be readily allowed. On the other hand, struture formation on
significantly smaller scales would be progressively suppressed, which would, in turn, 
be of relevance to the small scale problems associated with CDM structure formation.     

\section{Explicit forms of the density and its  gradient along parametric solution} 

To find $\left \lvert \frac{dr_i}{dr} \right \rvert$ 
we use the first of equations 
(\ref{eq:cycloid}), to obtain  
\begin{equation}
\frac{dr}{dr_i} = \frac{dA_{\rm sh}}{dr_i} \: (1 - \cos \theta) + 
A_{\rm sh} \: \sin \theta \: \frac{d\theta}{dr_i}. 
\end{equation}
This may be readily evaluated using 
$\frac{d\theta}{d\theta_i} \frac{d\theta_i}{dr_i}$,
and noting that, as we are studying divergences between 
neighboring cells at fixed time, the condition $dt =0$ may 
be imposed on the second of equations (\ref{eq:cycloid}), to obtain
\begin{equation}
    \frac{d\theta}{d\theta_i} = -\frac{\frac{dB_{\rm sh}}{d\theta_i}  (\theta - \sin \theta)}{B_{\rm sh}  (1 - \cos \theta )}.  
\end{equation}
For the family of profiles given by (\ref{eq:denprof}), and assuming 
an initial overdensity corresponding to the RMS 
fluctuations of the linear density 
field such that $\bar{\delta (R_i)} = \sigma_M (R_i)$
at the initial time $t_i$, one finds 
\begin{equation}
    \frac{d\theta_i}{dr_i} = \frac{\frac{10}{3} \: 2^{2/3} \: \sigma_M \: c \: \beta \: \left(\frac{r_i}{R_i}\right)^{\beta\ - 1}}{R_i \: \sin \theta}.  
\end{equation}
Also, as
$A_{\rm sh} = \frac{3}{10} \frac{r_i}{\bar{\delta}(r_i)}$ 
and $B_{\rm sh} = \frac{1}{2 H_0} \left (\frac{5}{3} \frac{\bar{\delta}(r_i)}{a_i}\right)^{-3/2}$
(Section~\ref{sec:model} and references therein), we have
\begin{eqnarray}
\frac{dA_{\rm sh}}{dr_i} &=& \frac{3}{10 \: \bar{\delta}(r_i)} + \frac{3}{10} 
\frac{2^{2/3} \: \sigma_M \: c \: \beta \: \left(\frac{r_i}{R_i} \right)^{\beta}}{\bar{\delta}(r_i)^2}; \\
\frac{dB_{\rm sh}}{d\theta_i} &=& -\frac{3 \: \sin \theta_i}{8 \: H_0 \: a_i}
\left (\frac{5 \: \bar{\delta}(r_i)}{3 \: a_i} \right )^{-5/2}.
\end{eqnarray}

The density evolution of the density of these profiles can then be evaluated along 
the parametric solution using equations (\ref{eq:simdens}) and (\ref{eq:voldens}).  

In order to obtain the density gradient,  
we need to evaluate  $\frac{d}{dr_i} \left \lvert \frac{dV_i}{dV} \right \rvert $. 
From (\ref{eq:voldens}), this first requires obtaining 
\begin{eqnarray}
\frac{d}{dr_i} \left(\frac{r_i^2}{r^2}\right) = \frac{200}{9} \: \frac{\bar{\delta}(r_i)} {(1 - \rm \cos \: \theta)^{2}}
\frac{d\bar{\delta}(r_i)}{dr_i}  \\ \nonumber 
-
\frac{200}{9} \: \bar{\delta}(r_i)^2  \: \frac{\sin \theta}{(1 - \rm \cos \theta)^{3}} \frac{d\theta}{dr_i}, \nonumber. 
\end{eqnarray}
In addition to 
\begin{equation}
 \frac{d}{dr_i} \left \lvert \frac{dr_i}{dr} \right \rvert = - \left \lvert \frac{dr}{dr_i} \right \rvert^{-2} \: \frac{d}{dr_i} \left \lvert \frac{dr}{dr_i} \right \rvert.
\end{equation}
The first derivative appearing here is obtained as described above. For the second derivative one has 
\begin{eqnarray}
\frac{d^2r}{dr_i^2} = 
&-&
\frac{3}{5 \: \bar{\delta}(r_i)^2} \: \frac{d\bar{\delta}(r_i)}{dr_i} \: (1 -\cos \theta) \\ \nonumber
&+&
\frac{3}{5 \: \bar{\delta}(r_i)} \: \sin \theta \: \frac{d\theta}{dr_i} \\ \nonumber
&+&
\frac{3}{10} \: \frac{2 \: r_i}{\bar{\delta}(r_i)^3} \: 
 \left (\frac{d\bar{\delta}(r_i)}{dr_i} \right )^2   \: (1 -\cos \theta) \\ \nonumber
&-&
\frac{3}{10} \: \frac{r_i}{\bar{\delta}(r_i)^2} \: 
 \frac{d^2\bar{\delta}(r_i)}{dr_i^2}  \: (1 -\cos \theta) \\ \nonumber
&-&
\frac{3}{5} \: \frac{r_i}{\bar{\delta}(r_i)^2} \: 
 \frac{d\bar{\delta}(r_i)}{dr_i} \: \sin \theta \: \frac{d\theta}{dr_i} \\ \nonumber
&+&
\frac{3}{10} \: \frac{r_i}{\bar{\delta}(r_i)} \: \cos \theta \left(\frac{d\theta}{dr_i} \right)^2 \\ \nonumber
&+&
\frac{3}{10} \: \frac{r_i}{\bar{\delta}(r_i)} \: \sin \theta \:  \frac{d^2\theta}{dr_i^2} \\ \nonumber.
\end{eqnarray}
Here
\begin{eqnarray}
\frac{d^2\theta}{dr_i^2} 
= 
&-&
\frac{10}{3} \: \frac{d\bar{\delta}(r_i)}{dr_i} \: \frac{\cos \theta_i}{\sin \: \theta_i^2} \: \frac{d\theta}{dr_i} \\ \nonumber
&+&  
\frac{10}{3} \: \frac{d^2\bar{\delta}(r_i)}{dr_i^2} \: \frac{1}{\sin \theta_i} \: \frac{d\theta}{d\theta_i} \\ \nonumber
&+&
\frac{9}{20} \: \frac{\cos \: \theta_i}{\bar{\delta}(r_i)} \:
\frac{\theta - \sin \theta}{1 - \cos \theta} \left ( \frac{d\theta_i}{dr_i}   \right)^2 \\ \nonumber
&-&
\frac{9}{20} \: \frac{1}{\bar{\delta}(r_i)^2} \:  \frac{d\bar{\delta}(r_i)}{dr_i} \: \sin \theta_i \:
\frac{\theta - \sin \theta}{1 - \cos \theta} \: 
 \frac{d\theta_i}{dr_i} \\ \nonumber
 &+&
\frac{9}{20} \: \frac{\sin \theta_i}{\bar{\delta}(r_i)} \:
 \frac{d\theta}{dr_i} \: \frac{d\theta_i}{dr_i} \\ \nonumber
 &-&
\frac{9}{20} \: \frac{\sin \theta_i}{\bar{\delta}(r_i)} \:
\frac{\sin \: \theta \left( \theta - \sin \theta \right) }{\left(1 - \cos \theta\right)^2} \: \frac{d\theta}{dr_i} \: \frac{d\theta_i}{dr_i} \\ \nonumber
\\ \nonumber
\end{eqnarray}

For the profiles (\ref{eq:denprof})  
\begin{equation}
 \frac{d\bar{\delta}(r_i)}{dr_i}   = - \sigma_M \: 2^{2/3} \: c \: \beta \: \frac{1}{R_i} \: \left(\frac{r_i}{R_i}\right)^{\beta - 1}
\end{equation}
and
\begin{equation}
\frac{d^2\bar{\delta}(r_i)}{dr_i^2}   = - \sigma_M \: 2^{2/3} \: c \: \: \frac{\beta \: \left(\beta-1\right)}{R_i^2} \: \left(\frac{r_i}{R_i}\right)^{\beta - 2}.
\end{equation}

\bibliography{Chaplygin}


\end{document}